%% file: main.tex
\documentclass[fleqn,usenatbib]{mnras}

\usepackage[T1]{fontenc}
\usepackage[varg]{newtx}
\usepackage{float}
\usepackage{physics}
\usepackage{subfiles}
\usepackage{comment}
\usepackage{booktabs}
\usepackage{graphics}
\usepackage{comment}
\usepackage{graphicx}
\usepackage{hyperref}
\usepackage{xcolor}
\usepackage{lipsum} 
\usepackage{tabularx}
\usepackage{graphicx}
\usepackage{orcidlink}
\DeclareRobustCommand{\VAN}[3]{#2}
\let\VANthebibliography\thebibliography
\def\thebibliography{\DeclareRobustCommand{\VAN}[3]{##3}\VANthebibliography}

\usepackage{graphicx}	
\usepackage{amsmath}	
\usepackage{xcolor}

\definecolor{blue_cust}{HTML}{0066ff}

\definecolor{gold_cust}{HTML}{FFBF00}

\definecolor{purple_cust}{HTML}{800080}

\definecolor{pink_cust}{HTML}{DC267F}

\definecolor{green_cust}{HTML}{00CC00}

\definecolor{perfect_green}{HTML}{4FBF26}

\definecolor{crimson}{HTML}{DC143C}

\newcommand{\grb}{AT2023vfi}
\newcommand{\gfo}{AT2017gfo}

\title[Luminosity predictions for kilonova emission]{Luminosity predictions for the first three ionisation stages of W, Pt and Au to probe potential sources of emission in kilonova}
\author[M. McCann et al]{
M. McCann$^{1}$\thanks{E-mail: michael.mccann@qub.ac.uk}\orcidlink{0000-0002-1532-1240}, L. P. Mulholland$^{1}$\orcidlink{0009-0003-2668-5589}, Z. 
Xiong$^{2}$\orcidlink{0000-0002-2385-6771}, C. A. Ramsbottom$^{1}$\orcidlink{0000-0003-1579-8556}, C. P. Ballance$^{1}$\orcidlink{0000-0003-1693-1793}, \newauthor \hspace{1.5mm}O. Just$^{2,3}$\orcidlink{0000-0002-3126-9913}, A. 
Bauswein$^{2,4}$\orcidlink{0000-0001-6798-3572}, G. 
Martínez-Pinedo$^{2,5,4}$\orcidlink{0000-0002-3825-0131}, F. McNeill$^{1}$\orcidlink{0009-0001-9528-7475}  and S. A. Sim$^{1}$\orcidlink{0000-0002-9774-1192} \\
$^{1}$Astrophysics Research Centre, School of Mathematics \& Physics, Queens University Belfast, BT7 1NN, Northern Ireland.\\
$^{2}$GSI Helmholtzzentrum f\"{u}r Schwerionenforschung, Planckstra{\ss}e 1, D-64291 Darmstadt, Germany.\\
$^{3}$Astrophysical Big Bang Laboratory, RIKEN Cluster for Pioneering Research, 2-1 Hirosawa, Wako, Saitama 351-0198, Japan.\\
$^{4}$Helmholtz Forschungsakademie Hessen f\"{u}r FAIR, GSI Helmholtzzentrum f\"{u}r Schwerionenforschung, Planckstra{\ss}e 1, 64291
Darmstadt, Germany.
\\
$^5$Institut f\"{u}r Kernphysik (Theoriezentrum), Fachbereich Physik,
Technische Universit\"{a}t Darmstadt, Schlossgartenstra{\ss}e 2, D-64289 Darmstadt, Germany.\\
}

\date{Accepted XXX. Received YYY; in original form ZZZ}
\pubyear{2024}

\begin{document}
\label{firstpage}
\pagerange{\pageref{firstpage}--\pageref{lastpage}}
\maketitle

\begin{abstract}
A large number of $R$-matrix calculations of electron impact excitation for heavy elements ($Z>70$) have been performed in recent years for applications in fusion and astrophysics research. With the expanding interest in heavy ions due to kilonova (KN) events such as \gfo~and \grb, this new data can be utilised for the diagnosis and study of observed KN spectra. In this work recently computed electron-impact excitation effective collision strengths are used, for the first three ionisation stages of tungsten (W, $Z=74$), platinum (Pt, $Z=78$) and gold (Au, $Z=79$), to construct basic collisional radiative models tailored for the late stage nebular phases of KN. Line luminosities are calculated at a range of electron temperatures and densities and the strengths of these lines for a representative ion mass are compared. For the case of W {\sc iii}, these optically thin intensities are additionally used to constrain the mass of this ion in both \gfo~and \grb. Comparing with theoretical predictions of nucleosynthesis yields from neutron-star merger simulations, broad agreement with the inferred ion masses of W is found. Furthermore, we highlight the value of W measurements by showing that the abundance of other groups of elements and outflow properties are constrained by exploiting theoretically motivated correlations between the abundance of W and that of lanthanides or third r-process peak elements. Based on simple estimates, we also show that constraints on the distribution of tungsten in the ejecta may be accessible through the line shape, which may also yield information on the neutron-star merger remnant evolution.
\end{abstract}


\section{Introduction}\label{sec:intro}

Since the observation of the neutron-star merger \gfo~kilonova (KN) event in 2017 \citep{smartt2017,pian2017}, several theoretical groups have been engaged in the possible identification of spectral features originating from r-process heavy elements. To date there has been confirmation of a P Cygni feature of Sr~{\sc ii} at approximately 1$\mu$m \citep{watson2019identification} and another P Cygni feature of  Y~{\sc ii} at 760nm \citep{sneppen2023discovery}. Collisional data using the $R$-matrix method for the corresponding transitions of these two features have since been calculated by \cite{mulholland2024_sr_y_ii}. In addition, a broad emission feature at 2.1$\mu$m has been suggested by \citep{hotokezaka2023tellurium,gillanders2024modelling} originating from a forbidden transition between the fine-structure levels of the ground configuration of Te {\sc iii}. This interpretation has been supported by NLTE modelling by \cite{hotokezaka2023tellurium} and more recently by \cite{Mulholland24Te}  using newly computed atomic data. A similar feature in \grb \;has also been investigated by \cite{levan2024heavy,Mulholland24Te,gillanders2024analysisjwstspectrakilonova}.

Other potential sources of the KN emission have been investigated by \cite{Hotokezaka2022WSe} when interpreting the observations of \gfo~by the Spitzer space telescope  \citep{Villar_2018_spitzer,Kasliwal22} in the late nebular phase at 43-74 days post merger. Strong emission was observed at 4.5$\mu$m and potential sources were listed as either Se {\sc iii} if the first r-process peak elements are abundant, or W {\sc iii} otherwise. Modelling this 4.5$\mu$m emission line was difficult due to the lack of accurate atomic data (i.e. wavelengths, Einstein A-coefficients and collisional electron-impact excitation rates) necessary for the NLTE modelling. In the analysis performed by \cite{Hotokezaka2022WSe}, calibrated wavelengths were obtained from the National Institute of Standards and Technology (NIST) database \citep{nist} and the Einstein A-coefficients were calculated using a formula derived by \cite{pasternack1940}, \cite{shortley1940} and \cite{bahcall1968} for M1 dipole transitions assuming LS coupling is valid.  The collisional atomic data were computed using the {\sc hullac} code \citep{barshalom2001} for transitions among the levels of the ground terms but otherwise a value of unity was assumed for the effective collision strengths. The authors themselves conclude that more accurate atomic data would be required to perform a more detailed analysis. The conclusions from this investigation found that the mid-infrared spectrum of \gfo~could match the synthetic spectrum at 4.5$\mu$m and the associated line would originate from Se {\sc iii} if the first r-process peak elements were abundant, but if the ejecta were dominated by elements beyond the first peak then W {\sc iii} would be a more realistic candidate with an associated W {\sc ii} line emerging at 6.05$\mu$m.

There are, however, more recent atomic data now available in the literature for the first three ionisation stages of W. In \cite{Smyth2018} Dirac $R$-matrix evaluations for the electron-impact excitation of W {\sc i} are provided, \cite{Dunleavy2022} compute similar quality $R$-matrix data for W {\sc ii} and more recently \cite{mccann2024electron} for W {\sc iii}. Using calibrated energy levels from NIST these three publications provide accurate Einstein A-values and effective collision strengths for a wide variety of incident electron temperatures. In this paper NLTE collisional radiative modelling is performed using these new atomic data sets for W {\sc i}, {\sc ii} and {\sc iii}, to compute photon luminosity predictions to further investigate the W {\sc iii} identification proposed by \cite{Hotokezaka2022WSe} for the 4.5$\mu$m forbidden line. The observability of the expected 6.05$\mu$m W {\sc ii} feature is also explored, which shows similar luminosities to the  4.5$\mu$m line when ion masses similar to that of W~{\sc iii} are incorporated into the model. Additionally, the mass estimates of W~{\sc iii} in \gfo~from \cite{Hotokezaka2022WSe} are revised using the newly calculated $R$-matrix data and a first mass estimate for W~{\sc iii} in \grb~is provided. The mass estimates for W~{\sc iii} are used to compare with theoretically predicted nucleosynthesis yields, the implications of the W measurements for the production of other heavy elements such as lanthanides, actinides and third-peak r-process elements can then be investigated. The mass estimate of W is also important to understand its production by the r-process. So far, it has been observed in only a few metal-poor stars~\citep{Roederer.Lawler.ea:2022}.

W is just one element located near the third r-process peak. Accurate $R$-matrix calculations have also been performed for the first three ionisation stages of Pt and Au, see \cite{Bromley2023} and \cite{Mccann2022} respectively. For completeness the photon luminosity calculations are repeated for both Pt and Au.

The outline of the paper is as follows. In Section \ref{sec:theory} the general theory of collisional radiative modelling and the calculation of optically thin luminosity emission spectra is reviewed. In Section \ref{sec:results} synthetic spectra at late-time KN relevant conditions are presented. In addition, representative luminosities are presented for the ten strongest lines of the first three ion stages of W, Pt and Au. Mass estimates of W {\sc iii} in both \gfo~and \grb~are made with the parameter space of electron density and temperature explored. Furthermore in Section~\ref{sec:hydro}, the implications of the mass measurements based on comparisons with hydrodynamical neutron-star merger simulations and nucleosynthesis calculations are discussed. Finally in Section \ref{sec:conc} the paper is concluded with a summary and outlook.

\section{Theory}\label{sec:theory}

For a specific ion, the population of a level $i$ relative to the ground state, $N_i$ can be determined from the collisional radiative equations \citep{bates1962recombination,Summers2006},
\begin{equation}
    \frac{dN_i}{dt} = \sum_j C_{ij} N_j ,
    \label{eq:creq}
\end{equation}
where the matrix $C_{ij}$ encompasses the rates of all the considered atomic processes connecting levels $i$ and $j$. The above formulation includes the rates of all atomic processes in the matrix $C_{ij}$. These include electron-impact excitation/de-excitation, spontaneous emission, ionization and recombination. At present, there is very limited data for the accurate modelling of ionisation (photoionisation or collisional) or recombination for the high-Z elements. Therefore, the analysis is restricted to electron-impact excitation/de-excitation and emission. These are likely to be the dominant processes for determining atomic level populations at late times. The timescales of these atomic processes are very fast compared to the expansion evolution of KN ejecta, therefore the populations of the different levels can be assumed to be in the steady state. This modelling allows for the estimation of the ion-masses in the ejecta but not necessarily elemental ones. 

The rates of spontaneous emission and electron-impact excitation and de-excitation have been calculated previously for tungsten (W, $Z=74$), platinum (Pt, $Z=78$) and gold (Au, $Z=79$) \citep{Smyth2018,Dunleavy2022,mccann2024electron,Bromley2023,Mccann2022,gillanders2021constraints}. The reader is referred to these respective articles for the specific details of the $R$-matrix calculations. Solving equation~\eqref{eq:creq} for the populations allows for estimations of emission in the steady state. The luminosities are calculated in terms of the photon-emissivity-coefficient (PEC), typically employed by the fusion community \citep{Summers2006} and defined by,
\begin{equation}
    \text{PEC}_{j\to i } = \frac{N_j A_{j \to i}}{n_e},
    \label{eq:pecs}
\end{equation}
where  $N_j$ is the upper level population normalised to the ground level, $A_{j \rightarrow i}$ is the Einstein A-coefficient for the transition from $j$ to $i$ and $n_e$ is the electron density. A PEC is a derived coefficient that is associated with a single spectral line and is often useful for predicting individual spectral line emission. The {\sc colradpy} package \citep{johnson2019colradpy} is employed to determine the populations and PECs. When calculating a luminosity to infer a specific ion-mass the populations must be re-normalised, this is due to the populations being calculated relative to the ground in equation~\eqref{eq:creq}. The populations have been re-normalised to be relative to the total population of the ion by including a factor of $\sum_i N_i$ when calulating the luminosity. The luminosities are therefore defined as,
\begin{equation}
    L_{j \to i} = \frac{hc}{\lambda_{j \to i}}   \frac{n_e\text{PEC}_{j\to i } }{\sum_i N_i} \frac{M_{\text{ion}}}{m_{\text{ion}}},
    \label{eq:lumo}
\end{equation}
in units of energy per time, where $hc/\lambda$ is the photon energy, $M_{\text{ion}}$ is the mass of the ion in the ejecta and $m_{\text{ion}}$ is the mass of a single ion particle. The ratio of $M_{\text{ion}}$ and $m_{\text{ion}}$ encodes the number of ions in the ejecta. By imposing some observed specific luminosity, $L_{j \to i}$ one can then make ion-mass estimates using equation~\eqref{eq:lumo}. Additionally, model ion spectra can be constructed and directly compared with observation.

\section{Results}\label{sec:results}

\begin{figure*}
    \centering
    \includegraphics[width=\textwidth]{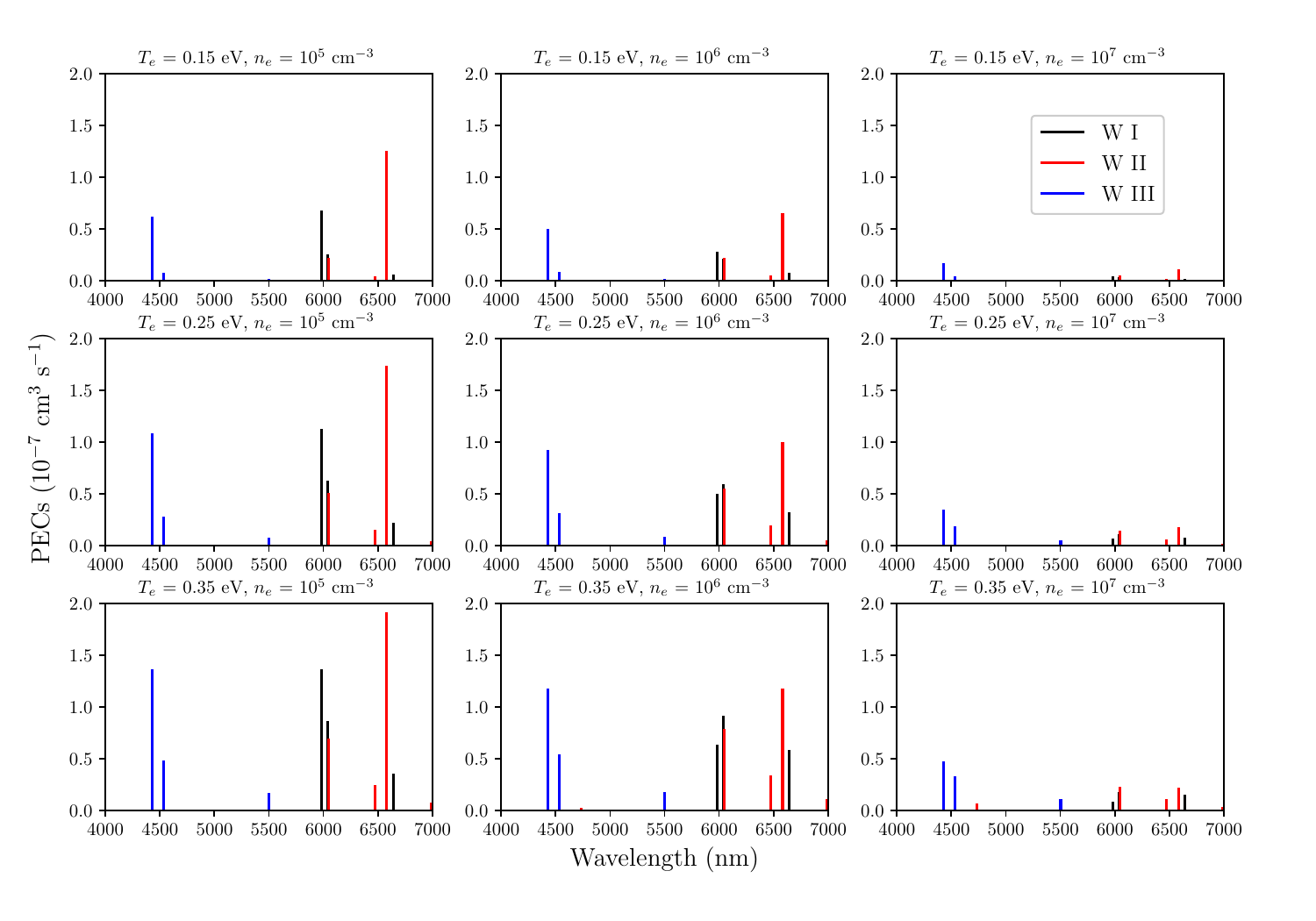}
    \caption{PECs for W~{\sc i}-{\sc iii} for a range of temperatures and densities ($T_e$ = 0.15, 0.25 and 0.35 eV, $n_e$ = 10$^5$, 10$^6$ and 10$^7$ cm$^{-3}$)}
    \label{fig:pecs_w}
\end{figure*}

\input{lumoW_2}

\subsection{Luminosities}\label{sec:luminosities}

\begin{figure*}
    \centering
    \includegraphics[width=0.9\textwidth]{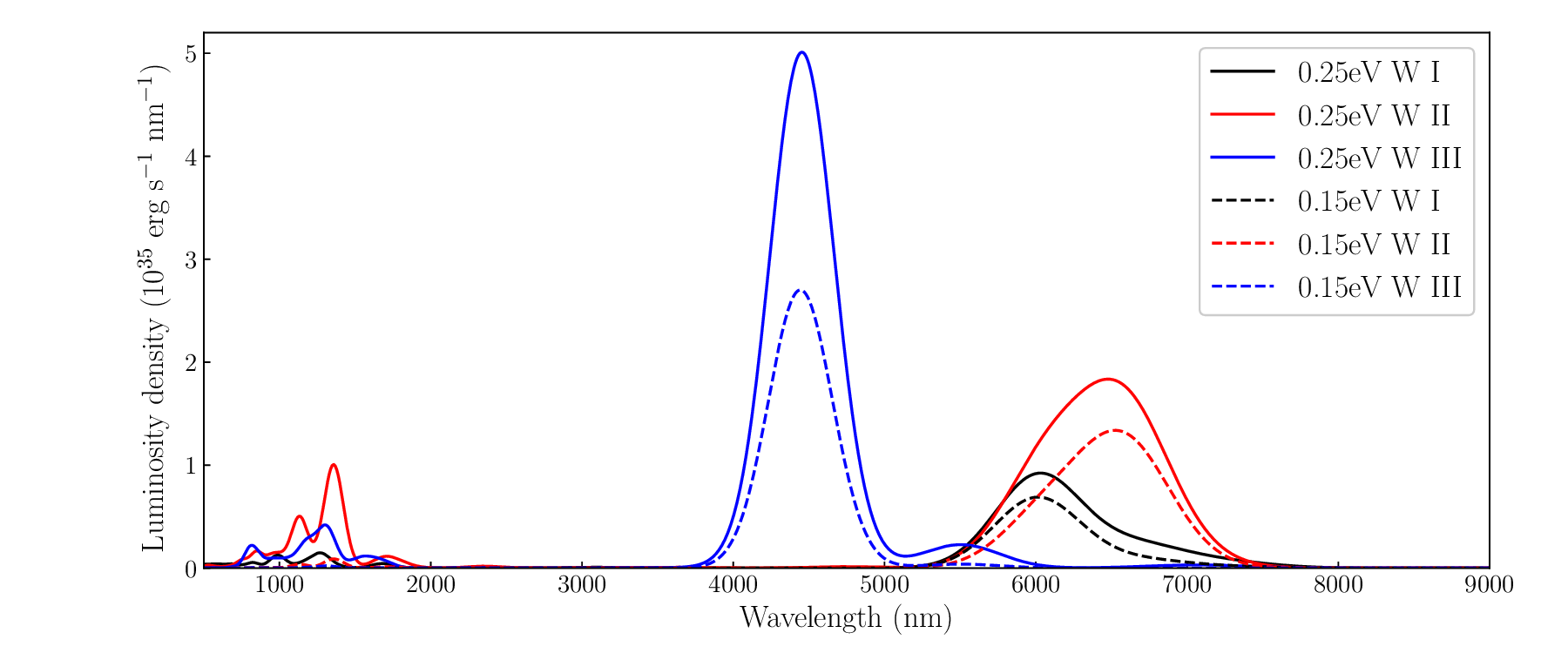}
    \caption{Luminosity density plot as a function of wavelength (nm) for W {\sc i} - {\sc iii} generated at $T_e = 0.15/0.25$~eV, $n_e = 1\times10^6$ cm$^{-3}$ and a mass of $1\times10^{-3} M_\odot $.}
    \label{fig:lumow}
\end{figure*}
\input{lumoAu_2}

\begin{figure*}
    \centering
    \includegraphics[width=0.9\textwidth]{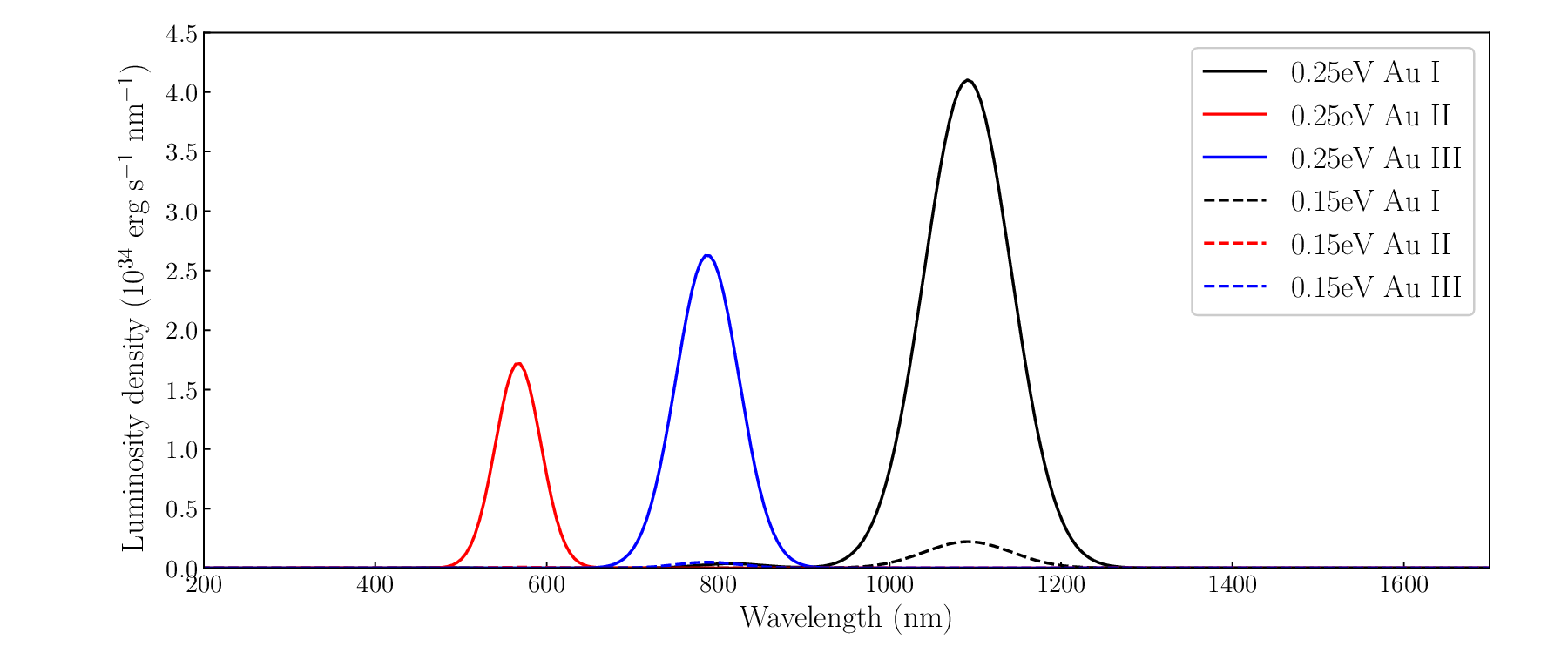}
    \caption{Luminosity density plot as a function of wavelength (nm) for Au {\sc i} - {\sc iii} generated at $T_e = 0.15/0.25$~eV, $n_e = 1\times10^6$ cm$^{-3}$ and a mass of $1\times10^{-3} M_\odot $.}
    \label{fig:lumoau}
\end{figure*}

\input{lumoPt_2}
\begin{figure*}
    \centering
    \includegraphics[width=0.9\textwidth]{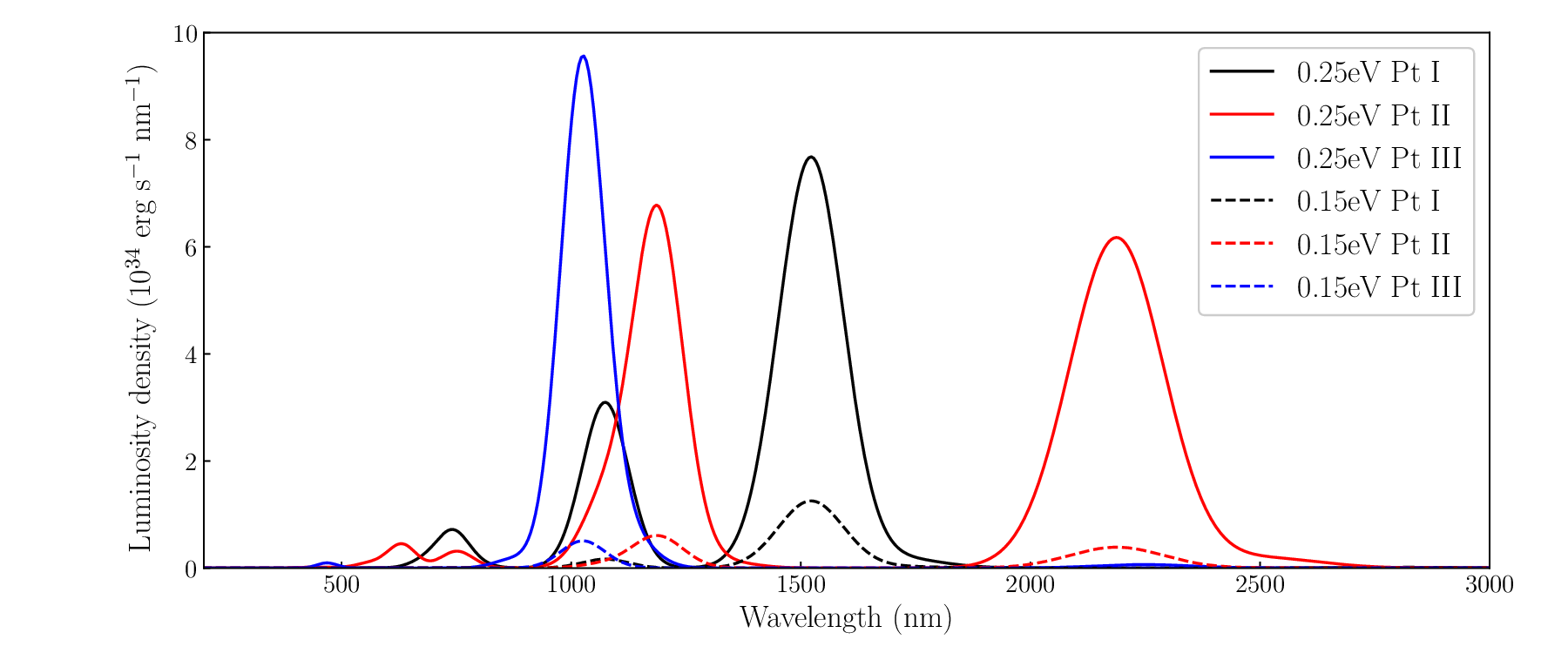}
    \caption{Luminosity density plot as a function of wavelength (nm) for Pt {\sc i} - {\sc iii} generated at $T_e = 0.15/0.25$~eV, $n_e = 1\times10^6$ cm$^{-3}$ and a mass of $1\times10^{-3} M_\odot $.
    }
    \label{fig:lumopt}
\end{figure*}

The electron density of the KN transient at late epochs is expected to be considerably lower than that of the early times. 
However, accurate values are difficult to estimate since they will depend on the radioactive heating, composition and ejecta density structure, which are hard to precisely constrain. Therefore, we will consider a range of electron densities in our analysis, $10^{5} - 10^{7}$cm$^{-3}$, which are in line with values adopted in previous studies \citep{hotokezaka2023tellurium,levan2024heavy}, for modelling \gfo~and \grb. Since these densities are below their typical critical-densities, M1 transitions among the fine-structure split terms of the ground configuration are expected to give rise to strong emission lines in the infrared region of the observed spectrum. All three charge states of W of interest in this publication have fine-structure transitions among their ground state, for W {\sc i} the ground level has terms 5d$^{4}$6s$^{2}$ $^{5}$D$_{0,1,2,3,4}$, for W {\sc ii} 5d$^{4}$6s $^{6}$D$_{1/2,3/2,5/2,7/2,9/2}$ and for W {\sc iii} 5d$^{4}$ $^{5}$D$_{0,1,2,3,4}$. W {\sc iii} is of particular interest as it has a strong fine-structure transition coincident with the  Spitzer space telescope 4.5$\mu$m wavelength range.

In this section the PECs  are investigated, defined in Section \ref{sec:theory} equation~\eqref{eq:pecs}, for all three ionization stages of W under consideration. In Figure~\ref{fig:pecs_w} these PEC coefficients are plotted (in units cm$^3$~s$^{-1}$) for W {\sc i}, W {\sc ii} and W {\sc iii}, as a function of wavelength (nm) from 4000 to 7000~nm spanning the spectrum from the near to far IR. Three electron temperatures are considered in the computations, $T_e$ = 0.15, 0.25 and 0.35 eV, and electron densities $n_e$ = 10$^5$, 10$^6$ and 10$^7$~cm$^{-3}$ of relevance to KN modelling. This covers the general temperature and density space used previously in the literature to studied specific spectral features \citep{Hotokezaka2022WSe,hotokezaka2023tellurium,levan2024heavy,gillanders2024modelling}. Clearly evident are three strong W {\sc i} lines (in black) at 5986.99, 6041.42 and 6646.86nm, representing the three low-lying forbidden transitions among the ground state terms 5d$^{4}$6s$^{2}$ $^{5}$D$^{\rm{o}}_{0}$ - 5d$^{4}$6s$^{2}$ $^{5}$D$^{\rm{o}}_{1}$, 5d$^{4}$6s$^{2}$ $^{5}$D$^{\rm{o}}_{1}$ - 5d$^{4}$6s$^{2}$ $^{5}$D$^{\rm{o}}_{3}$ and 5d$^{4}$6s$^{2}$ $^{5}$D$^{\rm{o}}_{3}$ - 5d$^{4}$6s$^{2}$ $^{5}$D$^{\rm{o}}_{4}$, respectively. These lines are evident in Figure \ref{fig:pecs_w} at all three temperatures considered and have their highest PEC values at the lower densities of 10$^{5}$ and 10$^{6}$ cm$^{-3}$. The peak lines of W {\sc ii} (in red) are found at 6047.25, 6477.50 and 6584.02nm and again represent forbidden transitions among the ground state terms, the 5d$^{4}$6s $^{6}$D$_{3/2}$ - 5d$^{4}$6s $^{6}$D$_{5/2}$, 5d$^{4}$6s $^{6}$D$^{\rm{o}}_{5/2}$ - 5d$^{4}$6s $^{6}$D$_{7/2}$ and 5d$^{4}$6s $^{6}$D$_{1/2}$ - 5d$^{4}$6s $^{6}$D$_{3/2}$ respectively. The 6041.42nm line of W {\sc i} and the 6047.25nm W {\sc ii} line are very close together in wavelength and could therefore become blended in physical spectra. Finally three strong W {\sc iii} lines (in blue) are predicted at 4432.23, 4535.16 and 5504.75nm, these correspond to the low lying forbidden lines 5d$^{4}$ $^{5}$D$_{0}$ - 5d$^{4}$ $^{5}$D$_{1}$, 5d$^{4}$ $^{5}$D$_{1}$ - 5d$^{4}$ $^{5}$D$_{2}$ and 5d$^{4}$ $^{5}$D$_{2}$ - 5d$^{4}$ $^{5}$D$_{3}$ respectively. Hence two potential W {\sc iii} lines are located around  4.5$\mu$m, where strong emission was detected by the Spitzer space telescope \citep{Kasliwal22}.

To quantify the relative strengths of these nine forbidden lines, Table~\ref{tab:lumow} gives the calculated luminosities in units ph s$^{-1}$ and erg s$^{-1}$ for the strongest ten lines of each W ion species. The photon luminosities are computed using equation~\eqref{eq:lumo} and adopting a reference mass of  $10^{-3}$ M$_\odot$, an electron temperature $T_{e}$ = 0.15 eV and an electron density of $n_{e}$ = 10$^{6}$ cm$^{-3}$. Clearly the strongest three lines for each species are those presented in the PEC plot of Figure~\ref{fig:pecs_w} and the largest luminosity value of 1.22 $\times$ 10$^{38}$ erg s$^{-1}$ is computed for the W {\sc iii} 4432.23nm line from the ground state to the first fine structure split level. It is likely due to large expansion velocities that this will blend with the slightly weaker 4535.16nm line ($\sim$ factor of 6 for 0.15~eV and 10$^{6}$ cm$^{-3}$), therefore both lines must be considered. This supports the W {\sc iii} prediction made by \cite{Hotokezaka2022WSe} for the 4.5$\mu$m forbidden line. These calculations also support the existence of a strong line at 6.05$\mu$m, the candidates for which could be either W {\sc i} or W {\sc ii}, due to blending in this wavelength region. This statement is subject to the ion mass of W {\sc i} or W {\sc ii} being comparable to W {\sc iii}, as it is for the calculations in Table~\ref{tab:lumow}.

In Figure~\ref{fig:lumow} a model spectrum of W {\sc i-iii} is plotted, which features line profiles broadened by a Gaussian kernel to produce a luminosity density spectrum. A full-width-half-maximum of 0.11$c$ \citep{gillanders2024analysisjwstspectrakilonova} was assumed for all lines. The kernels are normalised such that an integral over each line gives the total luminosity specified in Table~\ref{tab:lumow}. The spectrum is calculated at $n_e = 1.0 \times 10^6$ cm$^{-3}$ for two different electron temperatures roughly representative of nebular KN events. To investigate blending between the ions spectra, each ion is treated separately. Evidently the strongest feature is that of the two W {\sc iii} lines at 4.5~$\mu$m. Notably, at these broadening parameters these two lines present themselves as essentially a single Gaussian feature. At a similar mass, it is found again that the W {\sc i} and {\sc ii} lines at around 6.05~$\mu$m are of comparable strength. Additionally, there will be considerable blending between these two ion stages due to the overlap of the Gaussian profiles as seen in Figure~\ref{fig:lumow}. Nonetheless, it is feasible that future JWST observations could reach these large wavelengths in the IR, and in principle observe features here. This combined with the proposed W {\sc iii} identification may provide conclusive evidence of the element W in future KN like events.

Recent $R$ -matrix calculations by \cite{Mccann2022}, \cite{Bromley2023} and \cite{gillanders2021constraints} have provided new radiative and collisional excitation atomic data for the first three ionisation stages of Pt and Au. To check for any additional potential sources of KN emission from these species the photon luminosity calculations are repeated for Pt {\sc i} - {\sc iii} and Au {\sc i} - {\sc iii} using this new atomic data. In Table~\ref{tab:lumoau} the calculated luminosities in units ph s$^{-1}$ and erg s$^{-1}$ for the strongest ten lines of each Au ion species are presented, and in Table~\ref{tab:lumopt} the corresponding values for the three Pt ions. Again the photon luminosities are computed with a reference mass of  $10^{-3}~$M$_\odot$, an electron temperature $T_{e}$ = 0.15~eV and an electron density of $n_{e}$ = 10$^{6}$ cm$^{-3}$.~Clearly the luminosity strengths, for all six ions of both Au and Pt, are orders of magnitude lower than those predicted for W in Table~\ref{tab:lumow}, but the strongest transitions are similarly those among the low lying levels. For completeness in Figures \ref{fig:lumoau} and \ref{fig:lumopt} the corresponding broadened luminosity density profiles as a function of wavelength  are plotted, to visually display these strong features. Given the lack of discernible features at these wavelengths at the late times of \gfo~and \grb~for Pt  and Au, this places weak constraints on the masses of these elements as previously stated by \citep{gillanders2021constraints}. It is therefore the case that higher quality observed data is required to make more conclusive estimates of the presence of the elements in future KN events, in addition to accurate theoretical data for representative species.

\subsection{Ion-mass quantification for W}\label{sec:W_limits}

The late-time-Spitzer \citep{Villar_2018_spitzer,Kasliwal_spitzer} observation of  \gfo~and the observation of \grb~\citep{levan2024heavy} have produced interest in a potential feature at around $4.5 \mu$m. The ions {W {\sc iii}} and  {Se {\sc iii}}, which both have thermally accessible fine structure lines at around this wavelength. A first analysis was performed by \cite{Hotokezaka2022WSe} using distorted-wave collision data calculated using the {\sc hullac} code \citep{barshalom2001}. Here, a revised analysis of the parameter space for {W {\sc iii}} using the newly published $R$-matrix collisional data of \cite{mccann2024electron} for \gfo~is presented, and additionally a new analysis for \grb. This new atomic data is expected to be more accurate at low temperatures due to the inclusion of resonances when calculating the collision strengths.

For \gfo, the late time \emph{Spitzer} observation in the 4.5~$\mu$m band at 43 days had a total luminosity of $2 \times 10^{38}$ erg s$^{-1}$. It was suggested by \cite{Hotokezaka2022WSe} that W {\sc iii} could contribute as much as $5 \times 10^{37}$ erg s$^{-1}$ to this, where their model required a W {\sc iii} mass $\sim 2.0 \times 10^{-4}$ M$_{\odot}$, at a temperature of 3500~K and electron density of $10^6$ cm$^{-3}$. Enforcing these parameters and this integrated luminosity, the revised R-matrix data of \cite{mccann2024electron} requires a mass of $\sim 1.65 \times 10^{-4}$ M$_{\odot}$ of W {\sc iii}. Note that this estimate is based on the sum of the luminosities of both the ground configuration (5d$^{4}$6s$^{2}$) $^{5}$D$_{0}$ $\to $$^{5}$D$_{1}$ (4.43 $\mu$m) and $^{5}$D$_{1}$ $\to $$^{5}$D$_{2}$ (4.54 $\mu$m) transitions being $5 \times 10^{37}$ erg s$^{-1}$. In this case, the mass predicted is in agreement with that of \cite{Hotokezaka2022WSe}.  The good agreement between atomic data sets here shows the limiting factor in making such an estimate is likely the observed data, and in particular what fraction of the \emph{Spitzer} band to assign to this emission line. Additionally certain limitations are introduced with the assignment of an electron temperature and density. This estimate represents an ejecta mass  fraction of W {\sc iii} of around $0.33$\% in \gfo~(assuming an ejecta mass of $5\times10^{-2}$M$_{\odot}$ \cite{}), and assuming a similar production of other ion stages results in perhaps as much as $1.0$\% of the ejecta being W. While there is good agreement with the calculations of \cite{Hotokezaka2022WSe}, it is important to note that the methods here are more general. The use of a full collisional radiative model allows for mass predictions in any electron density regime, as opposed to assuming a coronal or LTE approximation. Additionally, including resonances when calculating the collision strengths allows for more accurate predictions at low temperatures - where  collision strengths could be underestimated using the distorted wave method.

For \grb, the feature has been integrated  to have a specific luminosity of $\sim1.0 \times 10^{38}$ erg s$^{-1}$ \citep{gillanders2024analysisjwstspectrakilonova}. Enforcing this luminosity, and a late time electron density of $\sim3 \times 10^5$~cm$^{-3}$ \citep{levan2024heavy}, at an electron temperature of 3000~K \citep{levan2024heavy}, the $R$-matrix data of \cite{mccann2024electron} requires a W {\sc iii} mass of $\sim9.4 \times 10^{-4}$ M$_{\odot}$. As for \gfo, this calculation is based on the sum of the 4.43 and 4.54 $\mu$m transitions. Given a total ejecta mass of $6 \times 10^{-2}$ M$_{\odot}$ \citep{levan2024heavy}, this mass of W {\sc iii} represents a fraction of around $1.6$\%. Without a quantitative treatment of ionization fraction, an elemental mass is difficult to quantify. Assuming reasonably similar fractions for the near neutral stages of W, it is perhaps feasible that up to $\sim4.7$\% could potentially be present in the ejecta. 

\begin{figure}
    \centering
    \includegraphics[width =\linewidth]{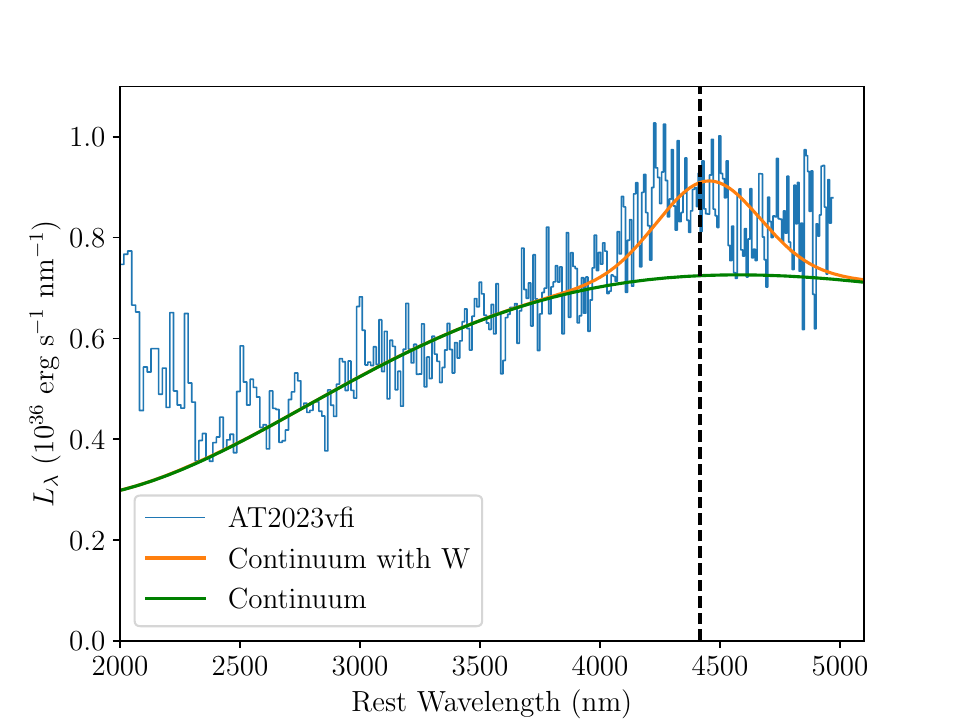}
    \caption{The calculated W {\sc iii} spectrum at $T_e = 0.26$~eV and $n_e = 3\times 10^5$~cm$^{-3}$, with full-width-half-maximum set to 0.110$c$ = 486~nm \citep{gillanders2024analysisjwstspectrakilonova}. A blackbody and powerlaw continuum is employed. This is overlaid on the JWST spectrum \citep{levan2024heavy}. The dashed black line shows the measured centroid of the line published by \citet{gillanders2024analysisjwstspectrakilonova}.}
    \label{fig:w2+grb}
\end{figure}

Figure \ref{fig:w2+grb} shows the synthetic W spectrum compared with the JWST spectrum of \grb~\citep{levan2024heavy}. The centroid and width of the $\sim 4.5 \mu$m feature was fitted by \cite{gillanders2024analysisjwstspectrakilonova}, which is also shown in Figure \ref{fig:w2+grb}. The spectrum was produced using an electron temperature of 0.26~eV and an electron density of $3\times 10^5$ cm$^{-3}$. A Gaussian kernel was again assumed with a full-width-half-maximum of $0.110c$ \citep{gillanders2024analysisjwstspectrakilonova}. With a W {\sc iii} mass of $9.4\times 10^{-4}$ M$_{\odot}$, a good match to the JWST data can be found assuming  a continuum of a power law added to a blackbody at 620K. This blackbody temperature is in rough agreement with those suggested by \citet{gillanders2024analysisjwstspectrakilonova} and \citet{levan2024heavy}.

\begin{figure}
    \centering
    \includegraphics[width =\linewidth]{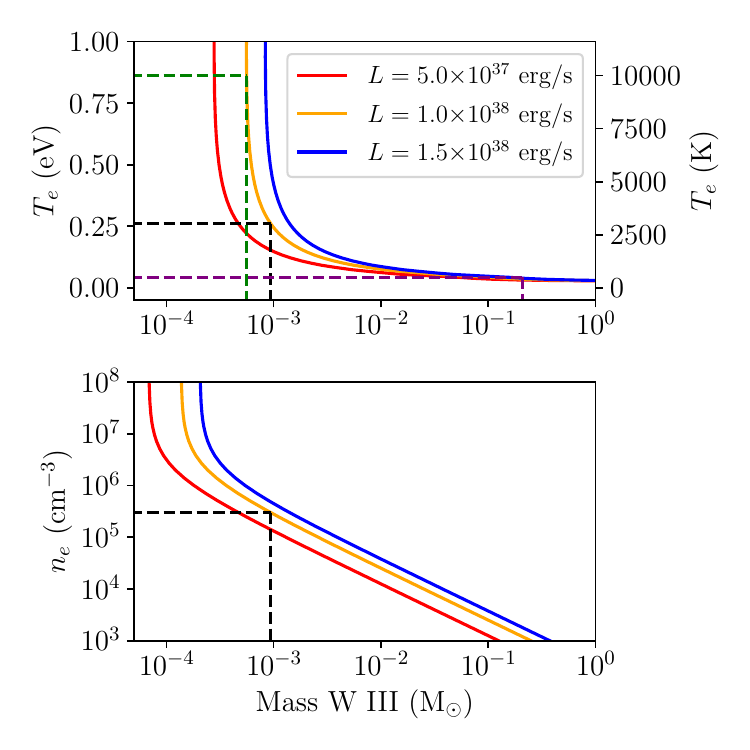}
    \caption{Contour plots of constant luminosity, at values of $L = 5.0\times 10^{37}$ erg/s, $L = 1.0\times 10^{38}$ erg/s and $L = 1.5\times 10^{38}$ erg/s. The parameter space of W {\sc iii} mass, electron temperature and electron density are explored. On the top panel, the density is fixed at 3$\times 10^{5}$ cm$^{-3}$. The dashed lines are fixed electron temperatures, namely 500K (0.04 eV, purple); 3,000K (0.26 eV, black) and 10,000K (0.86 eV, green).
    On the bottom panel, the electron temperature is fixed at 0.26 eV. The black dashed line is a constant density of 3$\times 10^{5}$ cm$^{-3}$.}
    \label{fig:w2+_contour}
\end{figure}

The non-uniformity of the electron density and temperature makes the assignment of either quantity in simple models like this difficult. For this reason an investigation is conducted across a wide range of temperature and density space to determine how the mass estimate varies subject to different conditions, in some sense accounting for a reasonable uncertainty in these quantities. This is shown in Figure~\ref{fig:w2+_contour} for a range of luminosities centered at $1.0 \times 10^{38}$ erg s$^{-1}$ to also investigate a rough uncertainty in the integration of the feature. At a fixed temperature the density and mass required to produce a fixed luminosity are inversely proportional in the limit of low density. This is simply a manifestation of the coronal regime of the collisional radiative equations. In contrast, the high density limit naturally requires less and less mass of W {\sc iii} to produce the features. 

\begin{figure*}
    \centering
    \includegraphics[width=0.9\textwidth]{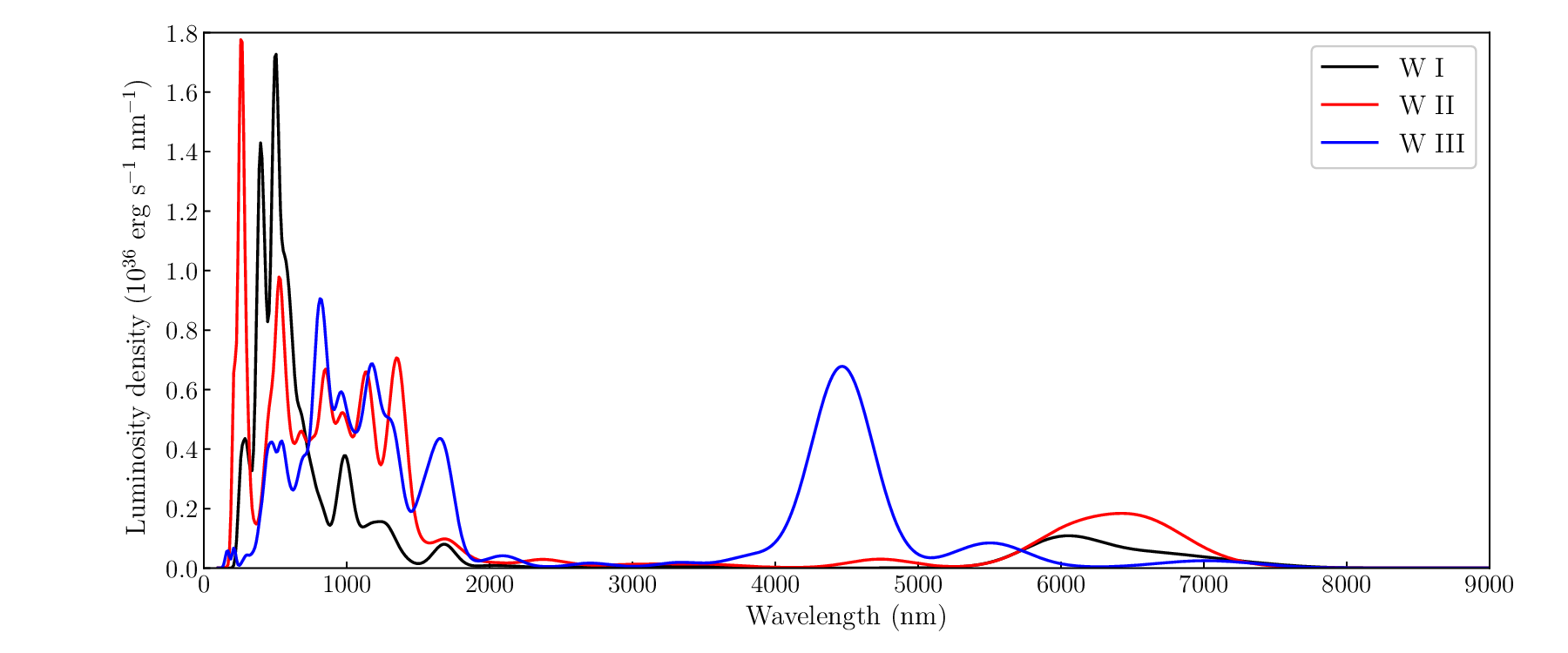}
    \caption{Luminosity density plot as a function of wavelength (nm) for W {\sc i} - {\sc iii} generated at $T_e = 0.86$~eV, $n_e = 1\times10^6$ cm$^{-3}$ and a mass of $1\times10^{-3} M_\odot $.}
    \label{fig:wiii_086eV}
\end{figure*}

With increasing temperature at fixed density it is naturally found that less mass of W {\sc iii} is required to produce the feature. The green dashed line on the top panel of Figure \ref{fig:w2+_contour} shows the implications of a temperature of 0.86 eV \cite[temperatures this high have been suggested by the models of][]{pognan_steady}. In this case, a relatively low mass of $5.6 \times 10^{-4}$ M$_{\odot}$ of W {\sc iii} is required to produce the feature. However, at such temperatures significant emission should begin to appear in the $\sim 1.0 - 1.8 \mu$m range from W {\sc ii} and {\sc iii} and at around $\sim 0.5 \mu$m from W {\sc i}, as shown in Figure~\ref{fig:wiii_086eV}. While this in principle could constrain the mass of W {\sc i}, the excess emission expected from W {\sc iii} that is not present in the observation perhaps prevents this from being a valid identification. In summary, for W {\sc iii} to be responsible for the excess flux at around $\sim 4.5 \mu$m at the rough density regime of $\sim 10^5$ cm$^{-3}$, it is required that the electron temperature also remain relatively low at around say 2500~K. Conversely, at low electron temperatures - say at around the $\sim 500$~K blackbody temperatures reported by \cite{gillanders2024analysisjwstspectrakilonova} and \cite{levan2024heavy}, the feature requires unreasonably high masses of around $2.1 \times 10^{-1}$ M$_{\odot}$ of W {\sc iii}. 

\begin{table*}
\begin{tabular}{rcccccccc}
\hline
Merger simulation& sym-n1-a6 & sym-n10-a3 & asy-n1-a6 & asy-n10-a3 & sym-n1-a6-short & asy-n1-a6-short & sym-n1-a6 & asy-n1-a6-short \\
Nuclear mass model& HFB21 & HFB21 & HFB21 & HFB21 & HFB21 & HFB21 & DZ31 & DZ31 \\ 
$\tau_{\mathrm{BH}}$\,[ms] & 122 & 915 & 96 & 680 & 10 & 10 & 122 & 10 \\\hline
$M_\mathrm{tot}\,[M_\odot]$ & 7.35(-2) & 3.25(-2) & 8.63(-2) & 6.13(-2) & 1.16(-2) & 2.44(-2) & 7.35(-2) & 2.44(-2) \\
$M$(W,1\,mth)$\,[M_\odot]$      & 5.57(-5) & 5.46(-5) & 6.96(-5) & 6.88(-5) & 4.51(-5) & 8.13(-5) & 1.63(-4) & 2.46(-4) \\
$M$(W,1\,Gyr)$\,[M_\odot]$      & 3.40(-5) & 3.30(-5) & 4.30(-5) & 4.19(-5) & 2.75(-5) & 4.99(-5) & 1.15(-4) & 1.77(-4) \\
$M(Y_e<0.2)\,[M_\odot]$     & 2.08(-3) & 2.05(-3) & 2.94(-3) & 3.11(-3) & 1.54(-3) & 3.74(-3) & 2.08(-3) & 3.74(-3) \\\hline
$X$(W,1\,mth)                             & 0.08\% & 0.17\% & 0.08\% & 0.11\% & 0.39\% & 0.33\% & 0.22\% & 1.01\% \\
$\frac{M(\mathrm{W,1\,mth})}{M(Y_e<0.2)}$ & 2.68\% & 2.66\% & 2.37\% & 2.21\% & 2.93\% & 2.17\% & 7.84\% & 6.58\% \\\hline
\end{tabular}
\caption{Ejecta masses and mass fractions in selected neutron-star merger models of \citealp{Just2023} and \citealp{Sneppen2024}. The second row lists the nuclear mass model employed for the nucleosynthesis post-processing of the ejecta, $\tau_{\mathrm{BH}}$ denotes the time after merger when the neutron-star remnant collapsed to a black hole, $M_\mathrm{tot}$ the total ejecta mass, $M$(W,1\,mth) and $M$(W,1\,Gyr) the total W masses one month and one billion years after the merger, respectively, $M(Y_e<0.2)$ the total amount of ejecta with electron fractions $Y_e$ below 0.2 at 5~GK, $X$(W,1\,mth) the mass fraction of W one month after merger, and the last row provides the ratio between the W mass at one month and ejecta mass with $Y_e<0.2$. Numbers are given with the notation $A(B):=A\times 10^{B} M_\odot$.}
\label{tab:XW}
\end{table*} 

\section{Implications of the limits on the ejected mass of W}\label{sec:hydro}

In this section, we compare theoretical merger models based on hydrodynamical simulations and nuclear network calculations with the estimated masses of W~{\sc iii} derived in the previous section, i.e. $1.65\times10^{-4}\,M_\odot$ for \gfo~ and $9.4 \times 10^{-4}\,M_\odot$ for \grb. Observational uncertainties of these estimates are discussed in the previous section, e.g. assumptions about the temperature and the possibility of line blending. The merger models only provide the total amount of W without distinguishing different ionization states and are themselves connected with uncertainties, some of which are briefly addressed below. Despite these uncertainties this comparison reveals a broad consistency between theoretical predictions and the observations and highlights the particular usefulness of W measurements to constrain the properties of the merger outflow and the abundance of other elements.

\begin{figure}
    \centering
    \includegraphics[width =\linewidth]{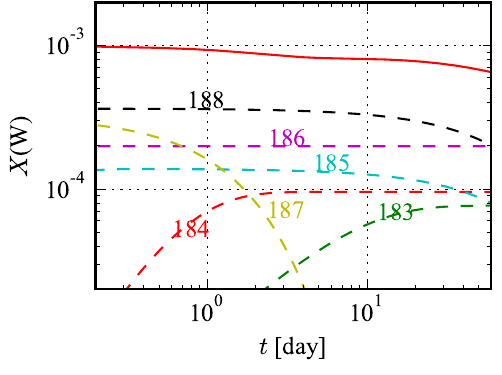}
    \caption{Time evolution of W mass fraction (solid red curve) from 0.2~day to 60~days averaged over all ejecta in the sym-n1-a6-HFB21 model. The contributions from different W isotopes are shown by dashed curves labeled by their mass numbers. The decays of $^{187}$W, $^{188}$W, and $^{185}$W lead to a moderate reduction of the total W abundance from $\sim 0.1$~day to $\sim 1$~month.}
    \label{fig:time_evolution_XW}
\end{figure}

In Table~\ref{tab:XW} we list the results from a selection of neutron-star merger models, drawn from \cite{Just2023} and \cite{Sneppen2024}, which strive for a complete and consistent description of {\it all } mass ejection channels.
The set consists of four models taken from \cite{Just2023}, two for equal-mass binaries ("sym") and two with a binary mass ratio of $q=0.75$ ("asy"). In order to capture the effects of turbulent angular-momentum transport during the post-merger evolution, these models adopt a viscosity model \citep{Shakura1973}. For each of the aforementioned two cases we consider one model with a fairly strong (``n1-a6'') and one model with relatively weak (``n10-a3'') viscosity. The models of \cite{Just2023} only involve relatively long-lived neutron-star remnants that collapse $\sim$100~ms after merger or later (cf. $\tau_\mathrm{BH}$ in Table~\ref{tab:XW}). We supplement these with two models (one equal mass, one asymmetric) from \cite{Sneppen2024} having short-lived remnants (i.e. collapsing at 10~ms after merger, labelled "short"; see references for details). The simulations consider neutron-stars with masses comparable to those in \gfo. The total amount of ejecta in these simulations is also broadly consistent with the estimated total ejecta mass of \gfo, namely about 0.03{--}0.05~$M_\odot$~\citep[e.g.][]{smartt2017}. 
In addition to the initial binary parameters and merger dynamics, nucleosynthesis yields are also affected by nuclear-physics uncertainties, which are particularly significant for heavy, neutron-rich nuclei. To explore this sensitivity, we add two further models that adopt the DZ31 nuclear mass model \citep{duflo1995microscopic} compared to the models in \cite{Just2023, Sneppen2024}, which adopted the HFB21 model \citep{Goriely2010} as well as the consistently derived $(n,\gamma)$ and $(\gamma,n)$ rates in \cite{mendoza2015nuclear}. These two models are chosen to have all other simulation parameters matching previous cases such that the influence of changing the nuclear physics input alone can be ascertained.

Elemental abundances in a KN are generally time dependent, because the radioactive decay half-lives are comparable to the relevant observable timescales of hours to months. The dominating decay mode, beta-decay, maintains constant the isotopic abundances but modifies the elemental abundances. To exemplify this, we show in Figure~\ref{fig:time_evolution_XW} the time evolution of the relative W mass fraction $X(W)$ resulting in the simulation ``sym-n1-a6'' with the HFB21 nuclear mass model along with the contributions from individual isotopes. At around 1 day, $^{187}$W decays, and the total mass fraction decreases. The decays of the other two isotopes $^{188}$W and $^{185}$W (with half lives of around 70 and 75~days, respectively) lead to a further reduction \cite[see supplemental material of][]{wu2019fingerprints}, despite a minor compensation from the decays of $^{184}$Ta and $^{183}$Ta. As a consequence, the mass fraction of W decreases by a factor of $\sim 2$ from the first hours to around two months after merger. Notice that this decrease should occur in any r-process model that reproduces the isotopic abundance pattern observed in the solar system around the third r-process peak.
Given that the observations of \gfo  ~and \grb  ~discussed in Section~\ref{sec:W_limits} were taken at 43 days and 29 days, respectively, after the merger, we base the following quantitative comparison on the mass fractions taken at 1 month, which are listed in Table~\ref{tab:XW}.

The predicted W masses scatter in the different merger models between $4 \times 10^{-5}$~M$_{\odot}$ and $2 \times 10^{-4}$~M$_{\odot}$ and are thus broadly consistent with the mass of W~{\sc iii} derived through the luminosity of the 4.5\,$\mu m$ feature in the spectra of \gfo~ and \grb~, providing tentative evidence that the inference presented in Section~\ref{sec:W_limits} yields reasonable estimates. 
The agreement is particularly good for \gfo. Although most of the models give W masses that are a factor of 2{--}3 smaller than our estimated $1.65 \times 10^{-4}$~M$_{\odot}$ (cf. Section~\ref{sec:W_limits}), the full spread within the models encompasses this value. Also, several of the simulations yield mass fractions for W, $X$(W), that are close to the estimated fraction of 0.3\,\% (Section~\ref{sec:W_limits}). The short-lived merger models tend to yield somewhat higher values of $X$(W) in slightly better agreement with the mass-fraction estimate from Section~\ref{sec:W_limits}, however, the total ejecta masses in these models are slightly too small to be compatible with \gfo. 
Considering \grb, the inferred W~{\sc iii} mass of $\sim 9.4 \times 10^{-4}\,$M$_\odot$ is up to one order of magnitude in excess of what the simulations predict, which were, however, originally set up to reproduce \gfo-like systems. At this point we cannot judge if this discrepancy is a deficiency of the approach adopted in this study keeping in mind, in particular, as noted in Section~\ref{sec:W_limits}, that the ion mass inferred from observations may be an overestimate if line blending is occurring, or only an upper limit if the feature is dominated by another process. 
Alternatively, this discrepancy may originate from different physical properties of the outflow in \grb~ compared to \gfo. As discussed below, a higher W mass points to more neutron-rich ejecta, which may result from a binary with different masses than \gfo. Since the gravitational-wave signal of \grb~was not detected, the binary masses in this event are unknown. High ejecta masses and very neutron-rich outflows may also result from a neutron-star black-hole merger \citep[e.g.][]{Kyutoku2021a}, which may also be a possible progenitor of \grb.
The numbers in Table~\ref{tab:XW} also show very clearly that the nuclear physics input for the network calculation has a significant impact on the W production, with DZ31 yielding systematically higher W masses (compare columns 2 and 8 or columns 7 and 9, respectively). Other aspects such as beta decay and fission rates may also impact the final abundance yields. The uncertainties of the nuclear model may thus represent another possible contribution explaining the difference between the theoretically predicted abundance and the W~{\sc iii} mass inferred from observations. Obviously, the current set of merger simulations is not exhaustive, and other physical ingredients (e.g. a different nuclear equation of state in the simulation) may yield different W masses. Moreover, due to their high level of complexity, the hydrodynamical simulations still adopt a number of approximations (e.g. in the neutrino transport and treatment of small-scale turbulence) and may not be fully resolved numerically.

\begin{figure}
    \centering
    \includegraphics[width =\linewidth]{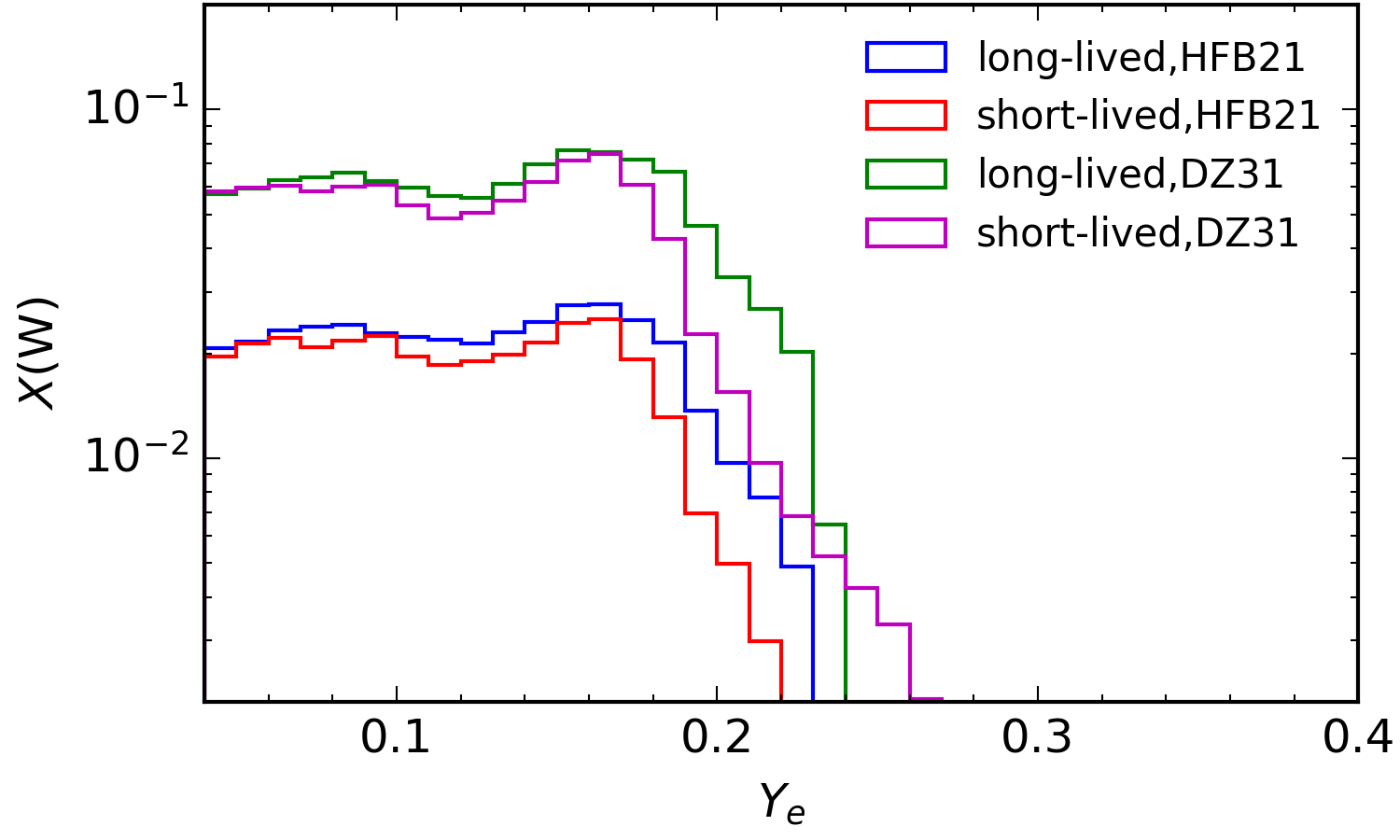}
    \caption{Histogram of W mass fraction measured at 1~month with respect to the electron fraction $Y_e$ for models sym-n1-a6 and asy-n1-a6-short with HFB21 and DZ31 nuclear mass inputs. This figure adopts the electron fraction of the individual tracer particles when they reach a temperature of 5~GK.}
    \label{fig:XW_ye}
\end{figure}

In general, the $r$-process is only sufficiently strong to produce significant amounts of W under neutron-rich conditions, i.e. if the electron fraction $Y_e$ characterizing the neutron-richness of the outflow is below 0.2--0.25. This is exemplified in Figure~\ref{fig:XW_ye}, where we bin the mass fraction of W as a function of $Y_e$ for the outflow tracer particles of four different merger models. Below $Y_e\approx0.2$ the W mass fraction is roughly constant in all models. The total W mass over the total ejected mass with $Y_e<0.2$ is given in Table \ref{tab:XW}, having similar values of $\sim 2.5\%$ for HFB21 models and $\sim 6$\% for DZ31 models.
The difference can be understood looking at the behaviour of the double neutron separation energy, $S_{2n}$ in the region of progenitor nuclei of stable W with $A\sim 186$. In the solar $r$-process abundance pattern W appears in the low-mass number tail of the 3rd peak. Its progenitors are located just before the neutron shell closure at $N=126$. 
In the DZ31 mass model, for an isotopic chain the $S_{2n}$ values decrease monotonically with respect to the mass number while they may become nearly constant or even increase in other models such as HFB21.
In the latter case, the competition between neutron capture and photodissociation reactions produces a trough in the abundance pattern in the region where the progenitor nuclei of W are located~\citep[see, e.g.,][]{arcones2011dynamical}.

The robust pattern for the ratio of the W mass to the total ejecta mass with $Y_e$ below 0.2, $M(Y_e<0.2)$, suggests the possibility to use a measured W mass as a proxy for the amount of neutron-rich ejecta, the theoretical prediction of which, however, is still affected by nuclear-physics uncertainties. Based on the values from Table~\ref{tab:XW}, neutron-rich material should roughly amount to about 40 times (14 times) the inferred W mass for HFB21 (DZ31), i.e. to $6.6\times 10^{-3}\,$M$_\odot$ ($2.3\times 10^{-3}\,$M$_\odot$) for \gfo~ and to $2.8\times 10^{-2}\,$M$_\odot$ ($9.8\times 10^{-3}\,$M$_\odot$) for \grb. For this neutron-rich material, the measured W mass constraints the production of non thermal electrons by beta decay. We find that for times later than a month  $^{188}$W represents the dominating contribution.

Potentially, a measured W mass fraction may provide a handle on the merger remnant life time since, within the set of simulations in Table~\ref{tab:XW}, $X$(W) is systematically reduced for long-lived models, for which a neutrino-driven wind increases the contribution of less neutron-rich material (e.g. \citealp{Perego2014a,Lippuner2017a,Fujibayashi2020b,Just2023}). This idea, however, requires more work to be solidified, as for instance model sym-n1-a6 with a low $X(\mathrm{W})$ has a shorter life time than sym-n10-a3.

\begin{figure}
    \centering
    \includegraphics[width =\columnwidth]{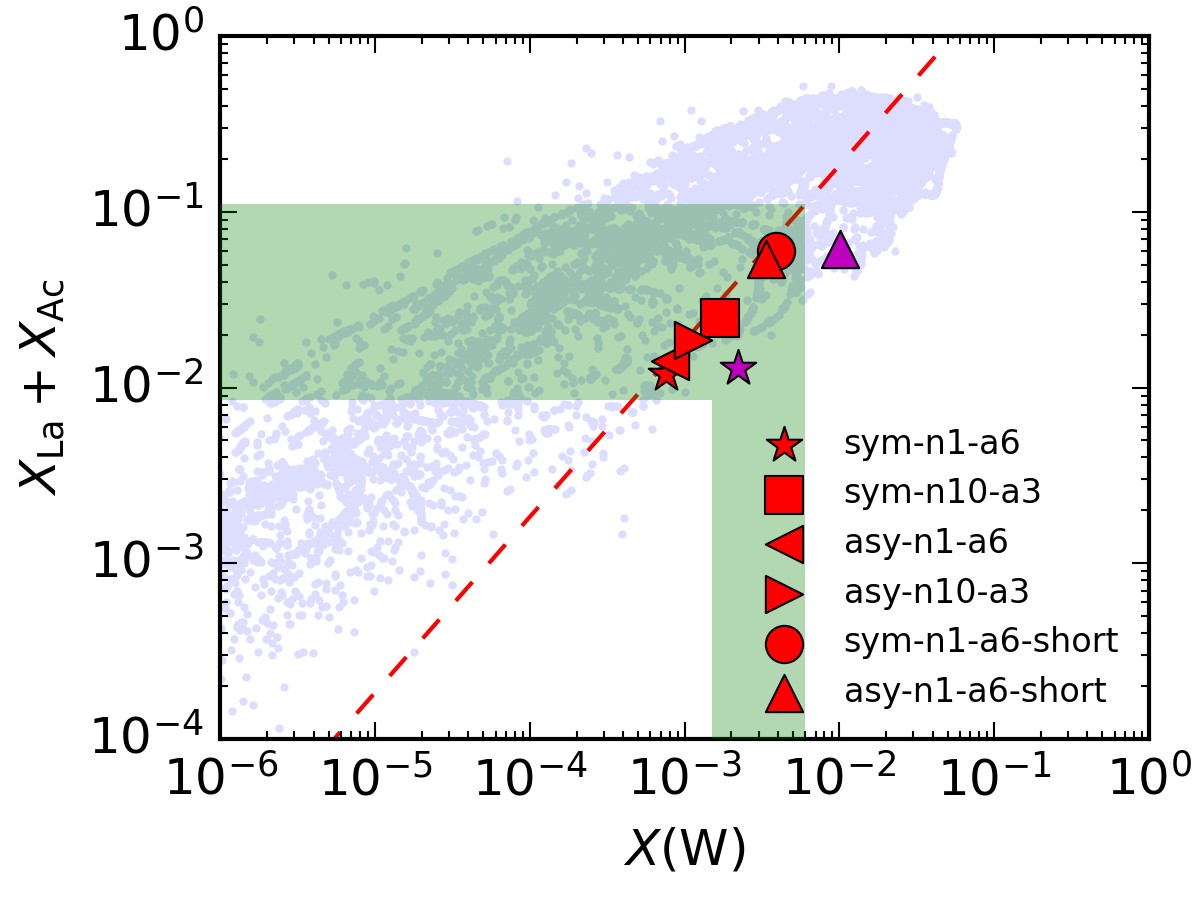}
    \includegraphics[width =\columnwidth]{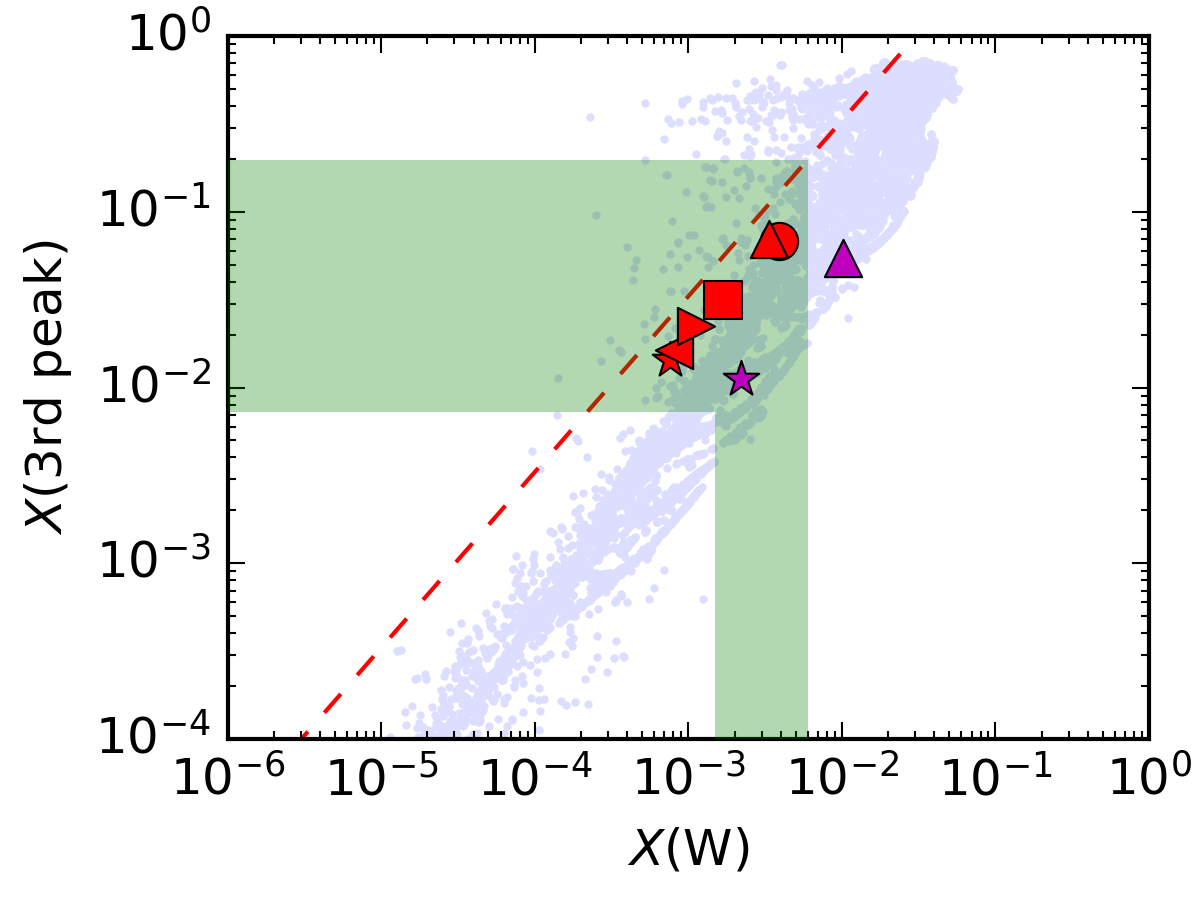}
    \caption{Correlation plots of W mass fraction with the total mass fraction of lanthanides and actinides, $X_\mathrm{La}+X_\mathrm{Ac}$, (upper panel) and with the mass fraction of third-peak $r$-process elements, $X$(3rd peak), (lower panel) for all models listed in Table~\ref{tab:XW} with red symbols for the HFB21 nuclear mass models and magenta markers for DZ31 based calculations. Red dashed lines display the ratio of mass fractions in solar $r$-process yields \citep{goriely1999uncertainties} with the W mass fraction converted to 1 month by including the presence of $^{185,188}$W that eventually decay into $^{185}$Re and $^{188}$Os (for the variation of W mass fractions between 1\,month and 1\,Gyr see Table~\ref{tab:XW}; we do not convert lanthanide and 3rd-peak mass fractions since the time dependence of averages over several elements and isotopes should be small). 
    The actinide mass fraction is subdominant compared to lanthanides at 1 month. Light blue dots show the mass fraction for all individual tracer particles of the HFB21 based models. The vertical green band indicates our observational estimate of $X$(W)$\approx 0.3\,\%$, with an estimated uncertainty of a factor of two towards both sides. The horizontal green bands cover the corresponding ranges adopting the DZ31 models (for the lower limit) and the solar abundance pattern (for the upper limit).}
    \label{fig:correlation_XW}
\end{figure}

Given the tight connection between the W mass and the mass of ejecta with $Y_e<0.2$ it is instructive to examine specific relationships between $X(\mathrm{W})$ and selected classes of species produced by low-$Y_e$ material. In particular, we address the extent to which $X(\mathrm{W})$ can be used as a tracer for the amount of third-peak elements or of lanthanides and actinides. 
Figure~\ref{fig:correlation_XW} shows the correlation of the W mass fraction with the mass fraction of all lanthanides and actinides, $X_\mathrm{La}+X_\mathrm{Ac}$, and the mass fraction of all 3rd $r$-process peak elements, $X$(3rd peak), respectively. Adopting the integrated yields of the individual simulations listed in Table~\ref{tab:XW}, the merger models for a given nuclear physics input (mass model) exhibit a very tight correlation implying that a measured mass fraction of W directly constrains the amount of lanthanides/actinides or of 3rd peak elements [as for instance gold and platinum]. Employing different nuclear physics input apparently yields a roughly constant offset of this correlation. The solar abundance pattern leads to a similar, but slightly shifted relation, reflecting the fact that the merger models do not exactly reproduce the solar composition~\citep[see][]{Just2023}.

A W mass fraction of 0.3\,\% as estimated in Section~\ref{sec:W_limits} corresponds to a mass fraction $1.8\%\lesssim X_\mathrm{La}+X_\mathrm{Ac}\lesssim 5.5\%$, where lanthanides dominate over actinides by far for the timescales considered here.
For the conversion we adopt the relation between $X(\mathrm{W})$ and  $X_\mathrm{La}+X_\mathrm{Ac}$ for the solar abundance to estimate the upper limit, and we employ the data points of the DZ31 models to determine a linear relation, which provides the lower bound. We assume W~{\sc iii} to be the dominant ionization state.

As an estimate of the uncertainties in the analysis and interpretation, which are typically a factor of a few (see Section~\ref{sec:W_limits}), in Figure~\ref{fig:correlation_XW} we show a green band corresponding to a factor of two variation in both directions and a horizontal green band illustrating the corresponding uncertainties in $X_\mathrm{La}+X_\mathrm{Ac}$ (using the same conversion as above). Despite the remaining uncertainties our study suggests that a significant amount of lanthanides was produced in \gfo~and \grb. We note that our estimated range of the lanthanide mass fraction (neglecting the small contributions from actinides) is in good agreement with the mass fraction of lanthanides that, according to the analysis of~\cite{Ji2019}, would be required by the observations of metal-poor stars for mergers to represent the main source of $r$-process elements. Interestingly, the high lanthanide mass fractions that we find exceed most of the estimates made based on the light curve of \gfo~ in the literature surveyed by \cite{Ji2019}. Similarly, our work predicts that a sizable amount of third-peak $r$-process elements of about $1.6\%\lesssim X(\mathrm{3rd~ peak})\lesssim 9.9\%$ was co-produced (see lower panel of Figure~\ref{fig:correlation_XW}).

Although we assume that the integrated yields of the different models and the solar composition represent a reasonable range for characterizing the integrated outflows of merger events like \gfo, we overplot $X_\mathrm{La}+X_\mathrm{Ac}$ and $X$(W) for all tracer particles of the six HFB21 models in Figure~\ref{fig:correlation_XW} (light blue dots) to visualize the local variation within the ejecta. The correlation between $X_\mathrm{La}+X_\mathrm{Ac}$ and $X$(W) is similar but not strictly linear with a larger spread of more than one order of magnitude in the mass fraction. 
As seen in Figure~\ref{fig:XW_ye} the mass fraction of W drops to zero when $Y_e$ changes from 0.18 to 0.25, i.e. in a very narrow range. The dots at low $X$(W) originate from this transition region and do not exactly follow the $X$(W) - ($X_\mathrm{La}+X_\mathrm{Ac}$) correlation favoring the production of lanthanides.
Visualizing the outcome of all tracers is likely overestimating the spread in the correlation because individual tracers may experience rather extreme conditions, which do not have a significant impact on the average behavior and are thus not representative of any viable full merger model.

\section{Velocity distribution}

\begin{figure}
    \centering
    \includegraphics[width =\linewidth]{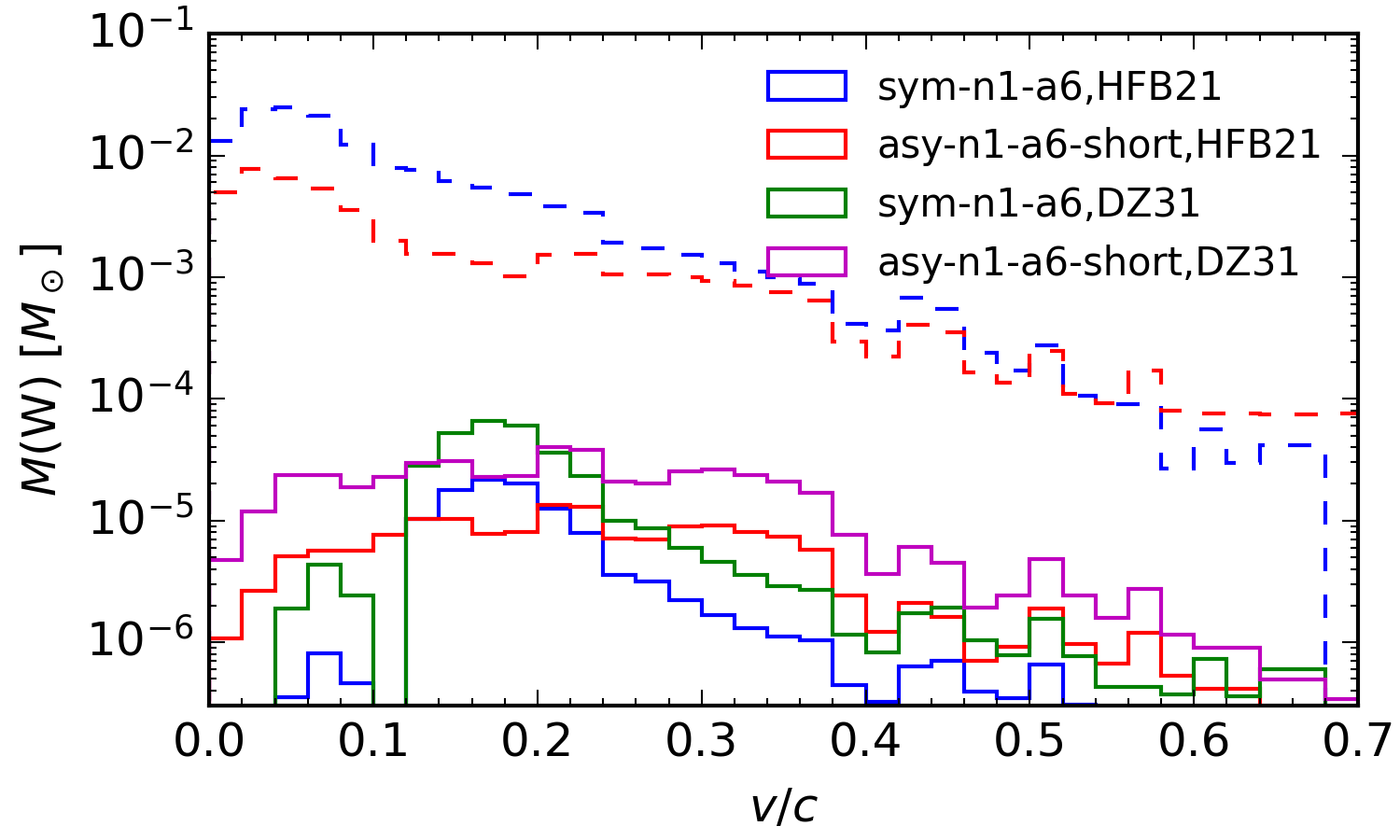}
    \caption{Histograms of $W$ mass (solid lines) as a function of radial velocity for models sym-n1-a6 and asy-n1-a6-short with HFB21 and DZ31 nuclear mass inputs (plotted for composition at 1 mth post merger). Dashed histograms show the total mass distributions, which are identical for our models that differ only in their nuclear mass inputs.}
    \label{fig:vr_W}
\end{figure}

A further test of consistency between merger models and observations is to consider the velocity distribution of the ejected $W$ in comparison to constraints on width of the observed spectral feature, which can be primarily attributed to Doppler broadening. Figure~\ref{fig:vr_W} shows the distribution of ejected $W$ mass versus radial velocity for four representative models from Table~\ref{tab:XW}. In all these models, the $W$ ejection velocity spans a fairly wide range, peaking around $\sim 0.2c$. There is a noticeable difference, however, in the amount of low velocity ($<0.1 c$) $W$ when comparing models with short-lived neutron-star remnants to those with long-lived remnants (the wind from a long-lived remnant leads to inner ejecta that are relatively $W$-poor).

\begin{figure}
    \centering
    \includegraphics[width =\linewidth]{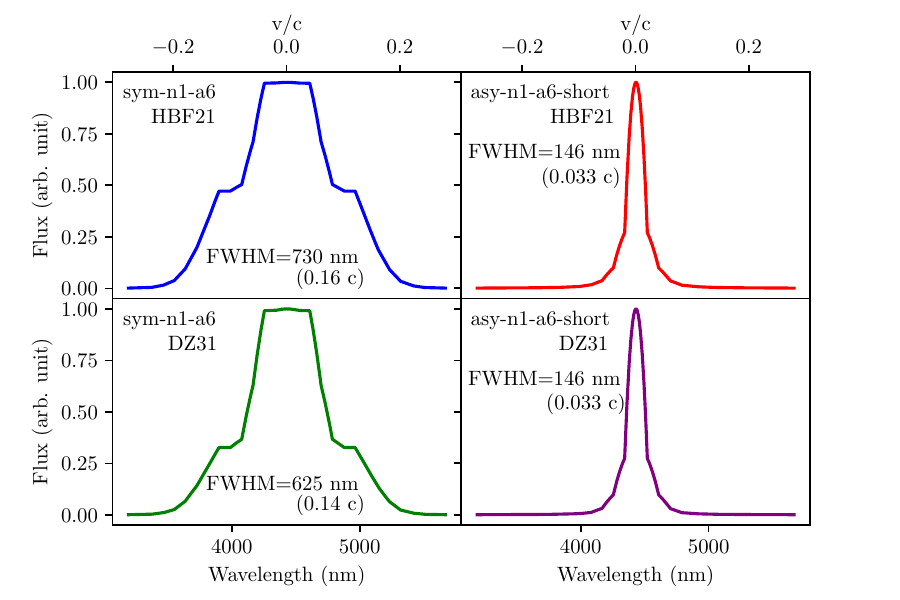}
    \caption{Line profile shapes computed from the $W$ mass and total mass distributions for models sym-n1-a6 and asy-n1-a6-short with HFB21 and DZ31 nuclear mass models. Each profile is plotted for the 4432.23~nm W~{\sc iii} line and normalised to peak. The full-width at have maximum (FWHM) is indicated for each case.}
    \label{fig:profiles}
\end{figure}

This difference in velocity distribution has implications for the expected profile shape which, given spectropscopic observations of sufficient quality, provides further constraints on the models. To illustrate this, we have computed simple optically-thin emission line profile shapes for the 4432.23~nm W~{\sc iii} transition (Figure~\ref{fig:profiles}) using the framework presented by \cite{Jerkstrand2017}, which is accurate to first order in $v/c$. For these calculations we do not take into consideration any temperature/ionization variations in the ejecta, or any departures from spherical symmetry, and thus they should be considered only illustrative, pending more detailed modelling. The calculations assume that the emission is optically thin and that the emissivity is proportional to the product of the $W$ particle density and the total mass density (adopted as a proxy for the free-electron density, under the assumption of uniform ionization).

Despite the simplicity of the calculations, these profiles indicate the potential to obtain information on the distribution of W in the ejecta, and possibly the fate of the merger remnant, through kilonova observations. In particular, our models with short-lived neutron-star remnants suggest narrow profiles, owing to the relatively high densities at low velocity. In contrast, the absence of $W$ at low velocities leads to broad, somewhat flat topped profiles for our models with long-lived neutron-star remnants. More sophisticated modelling is required to draw firm conclusions from comparisons with the observations of AT2023vfi, but these simple calculations suggest somewhat better compatibility of the short-lived remnant models with the relatively narrow observed profile (FWHM $\sim 0.11c$ \citealt[][]{gillanders2024analysisjwstspectrakilonova}), although we note that the present calculation yields profiles that are somewhat too narrow. This may be indicative of line blending or a more extended velocity distribution, but more detailed modelling (involving temperature and ionization calculations) is needed to draw firm conclusions.

\section{Conclusions}\label{conclusionsection}\label{sec:conc}

The available atomic data for W, Pt, and Au \citep{Smyth2018,Dunleavy2022,mccann2024electron,Bromley2023,Mccann2022} including transition rates and collision rates has been used to construct collisional-radiative models, producing level populations for these ions under KN conditions. This data has subsequently been used to make luminosity predictions for the aforementioned ions, see Tables~\ref{tab:lumow}, \ref{tab:lumoau} and \ref{tab:lumopt}. In this analysis it is evident that the luminosities of the features in Pt and Au are, in general, orders of magnitude lower than those of W, however this is when all the luminosities are calculated with the same $10^{-3} $M$_\odot$ mass. The blending of the features produced by these ions are also explored in Figures \ref{fig:lumow}, \ref{fig:lumoau} and \ref{fig:lumopt}, where the largest feature produced by the calculations at $4.5 \mu$m is a mix of the 4432.23nm and 4535.16nm lines from the 5d$^{4}$ $^{5}$D$_{0}$ - 5d$^{4}$ $^{5}$D$_{1}$ and 5d$^{4}$ $^{5}$D$_{1}$ - 5d$^{4}$ $^{5}$D$_{2}$ transitions in W {\sc iii}.

The suggested luminosities from observations of the \gfo~and the \grb, from \cite{Hotokezaka2022WSe} and \cite{gillanders2024analysisjwstspectrakilonova} respectively, have been used to determine mass estimates using the inverse procedure. Starting with the luminosity from observations, the required mass necessary to produce this value has been calculated using the available atomic data. For \gfo~ the measured luminosity was $5 \times 10^{37}$ erg s$^{-1}$, which resulted in a mass of $1.65 \times 10^{-4}$ M$_{\odot}$. For \grb~the measured luminosity was $1.0 \times 10^{38}$ erg s$^{-1}$, which resulted in a mass of $9.4 \times 10^{-4}$ M$_{\odot}$. A total ejecta mass of $5 \times 10^{-2}$ M$_{\odot}$ for \gfo~from \cite{Hotokezaka2022WSe} implies that $\sim 0.33\%$ of the ejecta mass is W~{\sc iii}, while for \grb~a total ejecta mass of $6 \times 10^{-2}$ M$_{\odot}$ from \cite{levan2024heavy} implies that $\sim 1.6\%$ of the ejecta mass is W~{\sc iii}. Assuming a similar amount of the first few ion stages this analysis would predict that $\sim 1.0\%$ of the \gfo~and $\sim 4.7\%$ of the \grb~ are from W.

To test plausibility the estimated amount of W~{\sc iii} is compared with the results from theoretical models of matter outflows of neutron-star mergers based on hydrodynamical simulations and nuclear-network calculations. These models, which resemble \gfo-like systems, yield total W masses in the range $4 \times 10^{-5}$~M$_{\odot}$ to $2 \times 10^{-4}$~M$_{\odot}$. Hence, broad consistency between the inferred masses from observations and theoretical predictions is found considering the involved uncertainties, which, on the theory side, include in particular the nuclear mass model employed in the nuclear network calculation. The mass determined from observation for \gfo~is close to the highest values predicted by our simulations, which is unsurprising due to the possibility that other lines or processes could be contributing to the feature in the observation. Concerning \grb, the comparison reveals a somewhat worse match which may either be related to uncertainties of the inference or point to the ejection of more neutron-rich material in \grb, for which no binary mass estimates exist since no gravitational waves were recorded from this event.

Nuclear network calculations suggest that the production of W requires generally very neutron-rich conditions with an electron fraction $Y_\mathrm{e} \le 0.2$, see Figure~\ref{fig:XW_ye}. Therefore, measurements of W are useful to estimate the amount of neutron-rich material from merger observations is therefore quite useful. Exploiting tight correlations between the W abundance and the mass fraction of lanthanides and actinides $X_\mathrm{La}+X_\mathrm{Ac}$, we convert the inferred mass fraction of W to an estimate of $X_\mathrm{La}+X_\mathrm{Ac}$. In the case of \gfo, we find mass fractions of these groups of elements in the range 0.85\,\% to 11\,\% of the ejecta mass. These elements featuring a very high opacity are particularly relevant for shaping the light curve of KN and estimates from the light curve suggested tentatively lower values, while observations for metal-poor stars point to a high production in mergers if those are the dominant site of r-process nucleosynthesis. Similarly, W can be used to estimate the amount of third-peak r-process elements, which are constrained to the range 0.73\% to 19\% for \gfo. It is therefore possible to infer from the observational estimate of the W mass in Section~\ref{sec:W_limits} that significant amounts of third-peak $r$-process elements, lanthanides and actinides will be produced along with it. Based on hydrodynamical merger models we estimate line profiles for the W feature, which are significantly affected by the tungsten production throughout the ejecta. This indicates the possibility to probe the velocity distribution of W and possibly distinguish short-lived and long-lived merger remnants if the latter indeed generically yield a strong deficiency of W at low velocities (v<0.1) as in our limited set of models. These considerations exemplify the usefulness of W to trace the properties of merger outflows and the underlying nucleosynthesis. It shows the importance of measuring W as laid out here and the need to improve observational data to more accurately determine the luminosity of features in KN spectra.

\section*{Funding}\label{sec:funding}

Funded/Co-funded by the European Union (ERC, HEAVYMETAL, 101071865). Views and opinions expressed are however those of the author(s) only and do not necessarily reflect those of the European Union or the European Research Council. Neither the European Union nor the granting authority can be held responsible for them. OJ, GMP and ZX acknowledge support by the European Research Council (ERC) under the European Union’s Horizon 2020 research and innovation programme (ERC Advanced Grant KILONOVA No. 885281).  AB, OJ, GMP, and ZX acknowledge support by the Deutsche Forschungsgemeinschaft (DFG, German Research Foundation) through Project - ID 279384907 – SFB 1245 (subprojects B01, B06, B07) and MA 4248/3-1. AB, OJ, GMP, and ZX acknowledge funding by the State of Hesse within the Cluster Project ELEMENTS. 

\section*{Acknowledgements}\label{sec:acknowledgements}

The authors are grateful for use of the computing resources from the Northern Ireland High Performance Computing (NI-HPC) service funded by EPSRC (EP/T022175).

\section*{Data Availability}
The atomic data used to calculate luminosities in this work will be made available on the \cite{openadas} database in the form of adf04 files.



\bibliographystyle{mnras}
\bibliography{main}




\appendix

\section{Effective Collision Strengths}

The following tables contain effective collision strengths for the ions that were discussed in Section~\ref{sec:luminosities} and include the lines shown in Tables~\ref{tab:lumow}, \ref{tab:lumoau} and \ref{tab:lumopt}. 
Tables~\ref{tab:eiew0}, \ref{tab:eiew1} and \ref{tab:eiew2} contain W~{\sc i}, W~{\sc ii} and W~{\sc iii}. Tables~\ref{tab:eiept0}, \ref{tab:eiept1} and \ref{tab:eiept2} contain Pt~{\sc i}, Pt~{\sc ii} and Pt~{\sc iii}. Tables~\ref{tab:eieau0}, \ref{tab:eieau1} and \ref{tab:eieau2} contain Au~{\sc i}, Au~{\sc ii} and Au~{\sc iii}.

\input{eietablew0}
\input{eietablew1}
\input{eietablew2}
\input{eietablept0}
\input{eietablept1}
\input{eietablept2}
\input{eietableau0}
\input{eietableau1}
\input{eietableau2}


\bsp	
\label{lastpage}
\end{document}

%% file: lumoW_2.tex
\begin{table*}
	\begin{tabular}{rcrrrrlcc}
		\hline
		\vspace{2mm}
		$\lambda$\phantom{00} &Index & $E_{{i}}$\phantom{00} & Lower & $E_{{j}}$\phantom{00} & Upper\phantom{-} & {$A_{{j\to i}}$} & \multicolumn{2}{c}{Luminosity $L$}  \\  
		(nm)&(${i}$-${j}$)\phantom{0} & (cm$^{-1}$) & ${i}$\phantom{00} &   (cm$^{-1}$) &  ${j}$\phantom{00} & (s$^{-1}$) & (ph s$^{-1}$) & (erg s$^{-1}$ ) \\
		\\
		\hline
		\\
		W {\sc i} lines &&&&&& \vspace{1mm}
		\\
		5986.99&      1 -  2&	     0.000&  5p$^6$5d$^4$6s$^2$ $^5$D$_{0}$&   1670.290& 5p$^6$5d$^4$6s$^2$ $^5$D$_{1}$&   5.32E-02&  8.41E+49&  2.79E+37 \\
		6041.42&      2 -  4&	  1670.290&  5p$^6$5d$^4$6s$^2$ $^5$D$_{1}$&   3325.530& 5p$^6$5d$^4$6s$^2$ $^5$D$_{2}$&   1.27E-01&  6.10E+49&  2.01E+37 \\
		6646.86&      4 -  5&	  3325.530&  5p$^6$5d$^4$6s$^2$ $^5$D$_{2}$&   4829.999& 5p$^6$5d$^4$6s$^2$ $^5$D$_{3}$&   1.35E-01&  2.11E+49&  6.31E+36 \\
		7197.71&      5 -  6&	  4829.999&  5p$^6$5d$^4$6s$^2$ $^5$D$_{3}$&   6219.330& 5p$^6$5d$^4$6s$^2$ $^5$D$_{4}$&   9.28E-02&  5.09E+48&  1.41E+36 \\
		1272.63&      2 -  7&	  1670.290&  5p$^6$5d$^4$6s$^2$ $^5$D$_{1}$&   9528.058& 5p$^6$5d$^4$6s$^2$ $^3$P$_{0}$&   8.07E+00&  1.14E+47&  1.79E+35 \\
		3059.94&      3 -  6&	  2951.289&  5p$^6$5d$^5$6s $^7$S$_{3}$&       6219.330& 5p$^6$5d$^4$6s$^2$ $^5$D$_{4}$&   2.33E-03&  1.28E+47&  8.31E+34 \\
		1682.76&      6 -  8&	  6219.330&  5p$^6$5d$^4$6s$^2$ $^5$D$_{4}$&  12161.954& 5p$^6$5d$^4$6s$^2$ $^3$H$_{4}$&   1.33E+00&  3.05E+46&  3.61E+34 \\
		1001.85&      4 -  9&	  3325.530&  5p$^6$5d$^4$6s$^2$ $^5$D$_{2}$&  13307.096& 5p$^6$5d$^4$6s$^2$ $^3$P$_{1}$&   5.72E+00&  8.45E+45&  1.68E+34 \\
		1173.91&      5 - 10&	  4829.999&  5p$^6$5d$^4$6s$^2$ $^5$D$_{3}$&  13348.555& 5p$^6$5d$^4$6s$^2$ $^3$G$_{3}$&   1.66E+00&  9.74E+45&  1.65E+34 \\
		1363.89&      5 -  8&	  4829.999&  5p$^6$5d$^4$6s$^2$ $^5$D$_{3}$&  12161.954& 5p$^6$5d$^4$6s$^2$ $^3$H$_{4}$&   3.45E-01&  7.92E+45&  1.16E+34 \\
		\\
		W {\sc ii} lines &&&&&& \vspace{1mm}
		\\
		6584.02&     1 -  2&	    0.000&  5p$^6$5d$^4$6s $^6$D$_{1/2}$&      1518.829& 5p$^6$5d$^4$6s $^6$D$_{3/2}$&     2.06E-01&  2.95E+50&  8.90E+37 \\
		6047.25&     2 -  3&	 1518.829&  5p$^6$5d$^4$6s $^6$D$_{3/2}$&      3172.472& 5p$^6$5d$^4$6s $^6$D$_{5/2}$&     2.80E-01&  1.00E+50&  3.29E+37 \\
		6477.50&     3 -  4&	 3172.472&  5p$^6$5d$^4$6s $^6$D$_{5/2}$&      4716.278& 5p$^6$5d$^4$6s $^6$D$_{7/2}$&     1.86E-01&  2.22E+49&  6.80E+36 \\
		6989.07&     4 -  5&	 4716.278&  5p$^6$5d$^4$6s $^6$D$_{7/2}$&      6147.084& 5p$^6$5d$^4$6s $^6$D$_{9/2}$&     8.60E-02&  3.80E+48&  1.08E+36 \\
		1367.26&     2 -  8&	 1518.829&  5p$^6$5d$^4$6s $^6$D$_{3/2}$&      8832.727& 5p$^6$5d$^4$6s $^4$P$_{1/2}$&     2.81E+00&  5.61E+47&  8.16E+35 \\
		1132.15&     1 -  8&	    0.000&  5p$^6$5d$^4$6s $^6$D$_{1/2}$&      8832.727& 5p$^6$5d$^4$6s $^4$P$_{1/2}$&     1.26E+00&  2.51E+47&  4.42E+35 \\
		1347.66&     1 -  6&	    0.000&  5p$^6$5d$^4$6s $^6$D$_{1/2}$&      7420.260& 5p$^6$5d$^5$ $^6$S$_{5/2}$&	   2.21E-02&  2.30E+47&  3.39E+35 \\
		1347.71&     3 -  9&	 3172.472&  5p$^6$5d$^4$6s $^6$D$_{5/2}$&     10592.484& 5p$^6$5d$^4$6s $^4$P$_{3/2}$&     7.06E-01&  1.68E+47&  2.48E+35 \\
		1694.50&     2 -  6&	 1518.829&  5p$^6$5d$^4$6s $^6$D$_{3/2}$&      7420.260& 5p$^6$5d$^5$ $^6$S$_{5/2}$&	   1.93E-02&  2.01E+47&  2.36E+35 \\
		1805.45&     3 -  7&	 3172.472&  5p$^6$5d$^4$6s $^6$D$_{5/2}$&      8711.273& 5p$^6$5d$^3$6s$^2$ $^4$F$_{3/2}$& 2.74E-02&  6.50E+46&  7.16E+34 \\
		\\
		W {\sc iii} lines &&&&&& \vspace{1mm}
		\\
		4432.23&      1 -  2&	    0.000&  5p$^6$5d$^4$ $^5$D$_{0}$&	    2256.199&	 5p$^6$5d$^4$ $^5$D$_{1}$&   5.04E-01&  2.73E+50&  1.22E+38 \\
		4535.16&      2 -  3&	 2256.199&  5p$^6$5d$^4$ $^5$D$_{1}$&	    4461.194&	 5p$^6$5d$^4$ $^5$D$_{2}$&   5.40E-01&  4.61E+49&  2.02E+37 \\
		5504.75&      3 -  4&	 4461.194&  5p$^6$5d$^4$ $^5$D$_{2}$&	    6277.808&	 5p$^6$5d$^4$ $^5$D$_{3}$&   2.60E-01&  6.93E+48&  2.50E+36 \\
		1307.51&      2 -  6&	 2256.199&  5p$^6$5d$^4$ $^5$D$_{1}$&	    9904.296&	 5p$^6$5d$^4$ $^3$P$_{0}$&   6.98E+00&  2.21E+47&  3.37E+35 \\
		7097.87&      4 -  5&	 6277.808&  5p$^6$5d$^4$ $^5$D$_{3}$&	    7686.681&	 5p$^6$5d$^4$ $^5$D$_{4}$&   7.13E-02&  9.06E+47&  2.54E+35 \\
		1536.73&      3 -  7&	 4461.194&  5p$^6$5d$^4$ $^5$D$_{2}$&	   10968.537&  5p$^6$5d$^3$6s $^5$F$_{1}$&   1.35E-01&  4.52E+46&  5.85E+34 \\
		1187.67&      3 -  9&	 4461.194&  5p$^6$5d$^4$ $^5$D$_{2}$&	   12881.030&	 5p$^6$5d$^4$ $^3$P$_{1}$&   3.60E+00&  3.16E+46&  5.29E+34 \\
		 804.69&      1 -  8&	    0.000&  5p$^6$5d$^4$ $^5$D$_{0}$&	   12427.091&  5p$^6$5d$^3$6s $^5$F$_{2}$&   9.31E-02&  1.13E+46&  2.78E+34 \\
		1662.71&      5 - 10&	 7686.681&  5p$^6$5d$^4$ $^5$D$_{4}$&	   13700.943&	 5p$^6$5d$^4$ $^3$H$_{4}$&   6.07E-01&  1.80E+46&  2.16E+34 \\
		1147.80&      2 -  7&	 2256.199&  5p$^6$5d$^4$ $^5$D$_{1}$&	   10968.537&  5p$^6$5d$^3$6s $^5$F$_{1}$&   3.00E-02&  1.00E+46&  1.74E+34 \\
		\\
		\hline
    \end{tabular}
	\caption{Luminosities for the ten strongest lines from each W ion calculated at $T_e = 0.15$ eV, $n_e = 1\times10^6$ cm$^{-3}$ and a mass of $1\times10^{-3} M_\odot $. The atomic data used in the calculation of luminosities for these ions is from \protect\cite{Smyth2018,Dunleavy2022,mccann2024electron}.}
	\label{tab:lumow}
\end{table*}

%% file: lumoAu_2.tex
\begin{table*}
	\begin{tabular}{rcrrrrlcc}
		\hline
		\vspace{2mm}
		$\lambda$\phantom{00} &Index & $E_{{i}}$\phantom{00} & Lower & $E_{{j}}$\phantom{00} & Upper\phantom{-} & {$A_{{j\to i}}$} & \multicolumn{2}{c}{Luminosity $L$}  \\  
		(nm)&(${i}$-${j}$)\phantom{0} & (cm$^{-1}$) & ${i}$\phantom{00} &   (cm$^{-1}$) &  ${j}$\phantom{00} & (s$^{-1}$) & (ph s$^{-1}$) & (erg s$^{-1}$ ) \\
		\\
		\hline
		\\
		Au {\sc i} lines &&&&&& \vspace{1mm}
		\\	
		1091.56&     1 -  2&      0.000&	5d$^{10}$6s $^2$S$_{1/2}$&	  9161.176&	5d$^9$6s$^2$ $^2$D$_{5/2}$&	2.48E-02&  1.56E+47&  2.84E+35 \\
		814.73&      2 -  3&   9161.176&	5d$^9$6s$^2$ $^2$D$_{5/2}$&	 21435.184&	5d$^9$6s$^2$ $^2$D$_{3/2}$&	2.98E+01&  1.76E+43&  4.30E+31 \\
		466.52&      1 -  3&      0.000&	5d$^{10}$6s $^2$S$_{1/2}$&	 21435.184&	5d$^9$6s$^2$ $^2$D$_{3/2}$&	1.31E+00&  7.75E+41&  3.30E+30 \\
		267.67&      1 -  4&      0.000&	5d$^{10}$6s $^2$S$_{1/2}$&	 37358.988&	5d$^{10}$6p $^2$P$_{1/2}$&	1.61E+08&  2.70E+37&  2.01E+26 \\
		242.87&      1 -  5&      0.000&	5d$^{10}$6s $^2$S$_{1/2}$&	 41174.609&	5d$^{10}$6p $^2$P$_{3/2}$&	2.26E+08&  1.29E+36&  1.06E+25 \\
		303.01&      2 -  6&   9161.176&	5d$^9$6s$^2$ $^2$D$_{5/2}$&      42163.529&	5d$^9$6s6p $^4$P$_{5/2}$&	1.37E+05&  7.50E+35&  4.92E+24 \\
		627.99&      3 -  4&  21435.184&	5d$^9$6s$^2$ $^2$D$_{3/2}$&      37358.988&	5d$^{10}$6p $^2$P$_{1/2}$&	2.33E+06&  3.91E+35&  1.24E+24 \\
		312.37&      2 -  5&   9161.176&	5d$^9$6s$^2$ $^2$D$_{5/2}$&      41174.609&	5d$^{10}$6p $^2$P$_{3/2}$&	2.60E+07&  1.49E+35&  9.46E+23 \\
		274.91&      2 -  7&   9161.176&	5d$^9$6s$^2$ $^2$D$_{5/2}$&      45537.183&	5d$^9$6s6p $^4$F$_{7/2}$&	1.43E+06&  8.34E+34&  6.04E+23 \\
		268.69&      2 -  9&   9161.176&	5d$^9$6s$^2$ $^2$D$_{5/2}$&      46379.043&	5d$^9$6s6p $^4$D$_{5/2}$&	1.21E+06&  1.89E+34&  1.40E+23 \\
		\\
		Au {\sc ii} lines &&&&&& \vspace{1mm}
		\\
		566.87&      1 -  3&	    0.000&	5p$^6$5d$^{10}$ $^1$S$_{0}$&	 17640.611& 5p$^6$5d$^9$6s $^3$D$_{2}$&       4.05E-01&   1.43E+45&   5.03E+33 \\
		3844.62&     2 -  3&	15039.574&	5p$^6$5d$^9$6s $^3$D$_{3}$&      17640.611& 5p$^6$5d$^9$6s $^3$D$_{2}$&       2.87E-01&   1.02E+45&   5.26E+32 \\
		987.64&      3 -  4&	17640.611&	5p$^6$5d$^9$6s $^3$D$_{2}$&      27765.754& 5p$^6$5d$^9$6s $^3$D$_{1}$&       2.72E+01&   1.97E+41&   3.96E+29 \\
		685.79&      2 -  5&	15039.574&	5p$^6$5d$^9$6s $^3$D$_{3}$&      29621.247& 5p$^6$5d$^9$6s $^1$D$_{2}$&       2.71E+01&   5.63E+40&   1.63E+29 \\
		337.60&      1 -  5&	    0.000&	5p$^6$5d$^{10}$ $^1$S$_{0}$&     29621.247& 5p$^6$5d$^9$6s $^1$D$_{2}$&       8.27E+00&   1.72E+40&   1.01E+29 \\
		834.68&      3 -  5&	17640.611&	5p$^6$5d$^9$6s $^3$D$_{2}$&      29621.247& 5p$^6$5d$^9$6s $^1$D$_{2}$&       1.74E+00&   3.61E+39&   8.61E+27 \\
		785.78&      2 -  4&	15039.574&	5p$^6$5d$^9$6s $^3$D$_{3}$&      27765.754& 5p$^6$5d$^9$6s $^3$D$_{1}$&       4.98E-03&   3.60E+37&   9.11E+25 \\
		5389.40&     4 -  5&	27765.754&	5p$^6$5d$^9$6s $^3$D$_{1}$&      29621.247& 5p$^6$5d$^9$6s $^1$D$_{2}$&       5.42E-02&   1.13E+38&   4.15E+25 \\
		393.09&      2 -  6&	15039.574&	5p$^6$5d$^9$6s $^3$D$_{3}$&      40478.743& 5p$^6$5d$^8$6s$^2$ $^3$F$_{4}$&   5.34E+00&   7.08E+36&   3.58E+25 \\
		437.86&      3 -  6&	17640.611&	5p$^6$5d$^9$6s $^3$D$_{2}$&      40478.743& 5p$^6$5d$^8$6s$^2$ $^3$F$_{4}$&   7.39E-01&   9.80E+35&   4.45E+24 \\
		\\
		Au {\sc iii} lines &&&&&& \vspace{1mm}
		\\
		787.77&      1 -  2& 	   0.000&    5d$^9$ $^2$D$_{5/2}$&    12694.038&    5d$^9$ $^2$D$_{3/2}$&     3.30E+01&   1.82E+46&   4.60E+34 \\
		336.09&      1 -  3& 	   0.000&    5d$^9$ $^2$D$_{5/2}$&    29753.996&    5d$^8$6s $^4$F$_{9/2}$&   1.34E-01&   7.04E+39&   4.17E+28 \\
		285.09&      1 -  4& 	   0.000&    5d$^9$ $^2$D$_{5/2}$&    35076.856&    5d$^8$6s $^4$F$_{7/2}$&   1.05E+01&   1.95E+38&   1.36E+27 \\
		1878.69&     3 -  4&   29753.996&  5d$^8$6s $^4$F$_{9/2}$&    35076.856&    5d$^8$6s $^4$F$_{7/2}$&   3.72E+00&   6.91E+37&   7.31E+25 \\
		257.58&      1 -  5& 	   0.000&    5d$^9$ $^2$D$_{5/2}$&    38822.353&    5d$^8$6s $^4$F$_{5/2}$&   7.95E+00&   4.56E+36&   3.52E+25 \\
		247.86&      1 -  6& 	   0.000&    5d$^9$ $^2$D$_{5/2}$&    40345.844&    5d$^8$6s $^4$F$_{3/2}$&   2.04E+01&   8.92E+35&   7.15E+24 \\
		382.73&      2 -  5&   12694.038&    5d$^9$ $^2$D$_{3/2}$&    38822.353&    5d$^8$6s $^4$F$_{5/2}$&   1.09E+00&   6.25E+35&   3.25E+24 \\
		446.77&      2 -  4&   12694.038&    5d$^9$ $^2$D$_{3/2}$&    35076.856&    5d$^8$6s $^4$F$_{7/2}$&   2.78E-02&   5.16E+35&   2.30E+24 \\
		2669.87&     4 -  5&   35076.856&  5d$^8$6s $^4$F$_{7/2}$&    38822.353&    5d$^8$6s $^4$F$_{5/2}$&   1.01E+00&   5.79E+35&   4.31E+23 \\
		225.09&      1 -  7& 	   0.000&    5d$^9$ $^2$D$_{5/2}$&    44425.917&    5d$^8$6s $^4$F$_{5/2}$&   1.72E+01&   1.86E+34&   1.65E+23 \\
		\\
		\hline
	\end{tabular}
	\caption{Luminosities for the ten strongest lines from each Au ion calculated at $T_e = 0.15$ eV, $n_e = 1\times10^6$ cm$^{-3}$ and a mass of $1\times10^{-3} M_\odot $. The atomic data used in the calculation of luminosities for these ions is from \protect\cite{Mccann2022}.}
	\label{tab:lumoau}
\end{table*}

%% file: lumoPt_2.tex
\begin{table*}
	\begin{tabular}{rcrrrrlcc}
		\hline
		\vspace{2mm}
		$\lambda$\phantom{00} &Index & $E_{{i}}$\phantom{00} & Lower & $E_{{j}}$\phantom{00} & Upper\phantom{-} & {$A_{{j\to i}}$} & \multicolumn{2}{c}{Luminosity $L$}  \\  
		(nm)&(${i}$-${j}$)\phantom{0} & (cm$^{-1}$) & ${i}$\phantom{00} &   (cm$^{-1}$) &  ${j}$\phantom{00} & (s$^{-1}$) & (ph s$^{-1}$) & (erg s$^{-1}$ ) \\
		\\
		\hline
		\\
		Pt {\sc i} lines &&&&&& \vspace{1mm}
		\\
		1522.66&      1 -  5&	       0.000&  5p$^6$5d$^9$6s $^3$D$_{3}$&       6567.450&  5p$^6$5d$^9$6s $^3$D$_{2}$&   	4.21E+00&   2.46E+48&   3.22E+36 \\
		12888.66&     1 -  2&	       0.000&  5p$^6$5d$^9$6s $^3$D$_{3}$&	  775.876&  5p$^6$5d$^9$6s $^1$D$_{2}$&   	3.26E-03&   3.64E+48&   5.62E+35 \\
		1076.07&      3 -  6&	     823.661&  5p$^6$5d$^8$6s$^2$ $^3$F$_{4}$&	10116.715&  5p$^6$5d$^8$6s$^2$ $^3$F$_{3}$&   	2.01E+01&   7.65E+46&   1.41E+35 \\
		1726.65&      2 -  5&	     775.876&  5p$^6$5d$^9$6s $^1$D$_{2}$&       6567.450&  5p$^6$5d$^9$6s $^3$D$_{2}$&   	1.13E-01&   6.61E+46&   7.62E+34 \\
		1068.83&      2 -  7&	     775.876&  5p$^6$5d$^9$6s $^1$D$_{2}$&      10131.867&  5p$^6$5d$^9$6s $^3$D$_{1}$&   	9.29E+00&   2.57E+46&   4.79E+34 \\
		1864.18&      2 -  4&	     775.876&  5p$^6$5d$^9$6s $^1$D$_{2}$&       6140.170&  5p$^6$5d$^{10}$ $^1$S$_{0}$&  	1.39E-02&   2.75E+46&   2.94E+34 \\
		740.95&       1 -  8&	       0.000&  5p$^6$5d$^9$6s $^3$D$_{3}$&      13496.261&  5p$^6$5d$^9$6s $^1$D$_{2}$&   	7.74E+00&   3.85E+45&   1.03E+34 \\
		1070.57&      2 -  6&	     775.876&  5p$^6$5d$^9$6s $^1$D$_{2}$&      10116.715&  5p$^6$5d$^8$6s$^2$ $^3$F$_{3}$&	9.81E-01&   3.73E+45&   6.94E+33 \\
		2805.51&      5 -  7&	    6567.450&  5p$^6$5d$^9$6s $^3$D$_{2}$&      10131.867&  5p$^6$5d$^9$6s $^3$D$_{1}$&   	1.07E+00&   2.96E+45&   2.10E+33 \\
		1741.01&      3 -  5&	     823.661&  5p$^6$5d$^8$6s$^2$ $^3$F$_{4}$&	 6567.450&  5p$^6$5d$^9$6s $^3$D$_{2}$&   	2.48E-03&   1.45E+45&   1.66E+33 \\
		\\
		Pt {\sc ii} lines &&&&&& \vspace{1mm}
		\\
		2188.35&     2 -  4&	    4786.652&   5p$^6$5d$^8$6s $^4$F$_{9/2}$&      9356.315& 5p$^6$5d$^8$6s1 $^4$F$_{7/2}$&   2.54E+00&   1.10E+48&   1.00E+36 \\
		1187.67&     1 -  3&	       0.000&	  5p$^6$5d$^9$ $^2$D$_{5/2}$&	   8419.839&	5p$^6$5d$^9$ $^2$D$_{3/2}$&   9.05E+00&   5.02E+47&   8.40E+35 \\
		1068.80&     1 -  4&	       0.000&	  5p$^6$5d$^9$ $^2$D$_{5/2}$&	   9356.315& 5p$^6$5d$^8$6s1 $^4$F$_{7/2}$&   1.10E-01&   4.77E+46&   8.87E+34 \\
		2517.02&     4 -  5&	    9356.315&   5p$^6$5d$^8$6s $^4$F$_{7/2}$&     13329.274& 5p$^6$5d$^8$6s1 $^4$F$_{5/2}$&   1.96E+00&   1.22E+46&   9.65E+33 \\
		633.26&      1 -  6&	       0.000&	  5p$^6$5d$^9$ $^2$D$_{5/2}$&     15791.307& 5p$^6$5d$^8$6s1 $^4$F$_{3/2}$&   3.85E+00&   8.20E+44&   2.57E+33 \\
		2036.89&     3 -  5&	    8419.839&	  5p$^6$5d$^9$ $^2$D$_{3/2}$&	  13329.274& 5p$^6$5d$^8$6s1 $^4$F$_{5/2}$&   2.68E-01&   1.67E+45&   1.63E+33 \\
		2089.14&     1 -  2&	       0.000&	  5p$^6$5d$^9$ $^2$D$_{5/2}$&	   4786.652& 5p$^6$5d$^8$6s1 $^4$F$_{9/5}$&   6.17E-06&   1.16E+45&   1.10E+33 \\
		1339.65&     4 -  7&	    9356.315&   5p$^6$5d$^8$6s $^4$F$_{7/2}$&     16820.928& 5p$^6$5d$^8$6s1 $^4$P$_{5/2}$&   6.46E+00&   6.38E+44&   9.47E+32 \\
		751.25&      2 -  8&	    4786.652&   5p$^6$5d$^8$6s $^4$F$_{9/2}$&     18097.766& 5p$^6$5d$^8$6s1 $^2$F$_{7/2}$&   1.52E+01&   3.48E+44&   9.20E+32 \\
		750.23&      1 -  5&	       0.000&	  5p$^6$5d$^9$ $^2$D$_{5/2}$&	  13329.274& 5p$^6$5d$^8$6s1 $^4$F$_{5/2}$&   3.93E-02&   2.45E+44&   6.49E+32 \\
		\\
		Pt {\sc iii} lines &&&&&& \vspace{1mm}
		\\
		1025.46&     1 -  3&	       0.000&	     5d$^8$ $^3$F$_{4}$&      9751.700&		5d$^8$ $^3$D$_{3}$&   4.66E+01&   3.15E+47&   6.10E+35 \\
		1126.28&     2 -  4&	    5293.100&	     5d$^8$ $^1$D$_{2}$&      14171.896&	5d$^8$ $^3$F$_{2}$&   2.22E+01&   5.22E+45&   9.22E+33 \\
		2242.86&     2 -  3&	    5293.100&	     5d$^8$ $^1$D$_{2}$&      9751.700&		5d$^8$ $^3$D$_{3}$&   1.13E+00&   7.63E+45&   6.76E+33 \\
		870.44&      2 -  6&	    5293.100&	     5d$^8$ $^1$D$_{2}$&      16781.599&	5d$^8$ $^3$P$_{1}$&   2.10E+01&   3.68E+44&   8.41E+32 \\
		2262.34&     3 -  4&	    9751.700&	     5d$^8$ $^3$D$_{3}$&      14171.896&	5d$^8$ $^3$F$_{2}$&   3.05E+00&   7.17E+44&   6.31E+32 \\
		468.81&      1 -  7&	       0.000&	     5d$^8$ $^3$F$_{4}$&      21330.798&	5d$^8$ $^1$G$_{4}$&   1.91E+01&   1.38E+43&   5.87E+31 \\
		3831.85&     4 -  6&	   14171.896&	     5d$^8$ $^3$F$_{2}$&      16781.599&	5d$^8$ $^3$P$_{1}$&   4.05E-01&   7.10E+42&   3.68E+30 \\
		863.63&      3 -  7&	    9751.700&	     5d$^8$ $^3$D$_{3}$&      21330.798&	5d$^8$ $^1$G$_{4}$&   1.94E+00&   1.41E+42&   3.24E+30 \\
		457.94&      1 -  8&	       0.000&	     5d$^8$ $^3$F$_{4}$&      21836.696&	5d$^7$6s $^5$F$_{5}$& 1.02E-02&	  3.94E+41&   1.71E+30 \\
		5427.12&     5 -  6&	   14938.999&	     5d$^8$ $^3$P$_{0}$&      16781.599&	5d$^8$ $^3$P$_{1}$&   2.00E-01&	  3.51E+42&   1.28E+30 \\
		\\
		\hline
	\end{tabular}
	\caption{Luminosities for the ten strongest lines from each Pt ion calculated at $T_e = 0.15$ eV, $n_e = 1\times10^6$ cm$^{-3}$ and a mass of $1\times10^{-3} M_\odot $. The atomic data used in the calculation of luminosities for these ions is from \protect\cite{Bromley2023}}
	\label{tab:lumopt}
\end{table*} 

%% file: eietableW0.tex
\begin{table*}
	\begin{tabular}{ccccccccc}
		\hline
Wavelength      &\multicolumn{2}{c}{Index}& A-value & \multicolumn{5}{c}{Effective Collision Strengths} \\ 
(nm) & Lower&Upper & (s$^{-1}$) & 0.15~eV & 0.25~eV & 0.4~eV & 0.6~eV & 0.8~eV \\
\hline
         5986.99&    1&    2&  5.32E-02&       1.03E+00&       1.02E+00&       1.01E+00&       1.01E+00&       1.00E+00\\ 
         6041.42&    2&    4&  1.27E-01&       2.25E+00&       2.29E+00&       2.33E+00&       2.37E+00&       2.39E+00\\ 
         6646.86&    4&    5&  1.35E-01&       3.47E+00&       3.47E+00&       3.47E+00&       3.47E+00&       3.47E+00\\ 
         7197.71&    5&    6&  9.28E-02&       3.64E+00&       3.70E+00&       3.75E+00&       3.80E+00&       3.83E+00\\ 
         1272.63&    2&    7&  8.07E+00&       3.42E-01&       3.29E-01&       3.17E-01&       3.06E-01&       2.99E-01\\ 
         3059.94&    3&    6&  2.33E-03&       3.25E+00&       3.16E+00&       3.08E+00&       3.01E+00&       2.96E+00\\ 
         1682.76&    6&    8&  1.33E+00&       1.86E+00&       1.68E+00&       1.51E+00&       1.36E+00&       1.26E+00\\ 
         1001.85&    4&    9&  5.72E+00&       3.15E-01&       3.08E-01&       3.02E-01&       2.97E-01&       2.93E-01\\ 
         1173.91&    5&   10&  1.66E+00&       9.83E-01&       9.02E-01&       8.27E-01&       7.63E-01&       7.17E-01\\ 
         1363.89&    5&    8&  3.45E-01&       8.80E-01&       7.95E-01&       7.17E-01&       6.50E-01&       6.02E-01\\ 
        \hline
    \end{tabular}
	\caption{Effective collision strengths for W~{\sc i}.}
	\label{tab:eiew0}
\end{table*}

%% file: eietableW1.tex
\begin{table*}
	\begin{tabular}{ccccccccc}
		\hline
Wavelength      &\multicolumn{2}{c}{Index}& A-value & \multicolumn{5}{c}{Effective Collision Strengths} \\ 
(nm) & Lower&Upper & (s$^{-1}$) & 0.15~eV & 0.25~eV & 0.4~eV & 0.6~eV & 0.8~eV \\
\hline
         6584.02&    1&    2&  2.06E-01&       4.07E+00&       3.76E+00&       3.38E+00&       2.99E+00&       2.74E+00\\ 
         6047.25&    2&    3&  2.80E-01&       5.41E+00&       5.10E+00&       4.69E+00&       4.25E+00&       3.94E+00\\ 
         6477.50&    3&    4&  1.86E-01&       7.39E+00&       6.91E+00&       6.33E+00&       5.77E+00&       5.41E+00\\ 
         6989.07&    4&    5&  8.60E-02&       7.99E+00&       7.61E+00&       7.11E+00&       6.53E+00&       6.10E+00\\ 
         1367.26&    2&    8&  2.81E+00&       8.95E-01&       8.50E-01&       7.82E-01&       6.99E-01&       6.44E-01\\ 
         1132.15&    1&    8&  1.26E+00&       7.20E-01&       6.58E-01&       5.71E-01&       4.82E-01&       4.26E-01\\ 
         1347.66&    1&    6&  2.21E-02&       7.95E-01&       7.72E-01&       7.18E-01&       6.49E-01&       6.04E-01\\ 
         1347.71&    3&    9&  7.06E-01&       1.96E+00&       1.84E+00&       1.65E+00&       1.45E+00&       1.31E+00\\ 
         1694.50&    2&    6&  1.93E-02&       2.01E+00&       1.95E+00&       1.84E+00&       1.69E+00&       1.60E+00\\ 
         1805.45&    3&    7&  2.74E-02&       1.88E+00&       1.86E+00&       1.75E+00&       1.57E+00&       1.44E+00\\ 
        \hline
    \end{tabular}
	\caption{Effective collision strengths for W~{\sc ii}.}
	\label{tab:eiew1}
\end{table*}

%% file: eietableW2.tex
\begin{table*}
	\begin{tabular}{ccccccccc}
		\hline
Wavelength      &\multicolumn{2}{c}{Index}& A-value & \multicolumn{5}{c}{Effective Collision Strengths} \\ 
(nm) & Lower&Upper & (s$^{-1}$) & 0.15~eV & 0.25~eV & 0.4~eV & 0.6~eV & 0.8~eV \\
\hline
         4432.23&    1&    2&  5.04E-01&       1.74E+00&       1.58E+00&       1.43E+00&       1.30E+00&       1.21E+00\\ 
         4535.16&    2&    3&  5.40E-01&       4.59E+00&       4.11E+00&       3.67E+00&       3.29E+00&       3.02E+00\\ 
         5504.75&    3&    4&  2.60E-01&       5.79E+00&       5.28E+00&       4.81E+00&       4.41E+00&       4.12E+00\\ 
         1307.51&    2&    6&  6.98E+00&       6.23E-01&       5.50E-01&       4.83E-01&       4.25E-01&       3.84E-01\\ 
         7097.87&    4&    5&  7.13E-02&       8.00E+00&       7.21E+00&       6.48E+00&       5.84E+00&       5.40E+00\\ 
         1536.73&    3&    7&  1.35E-01&       1.81E+00&       1.63E+00&       1.47E+00&       1.33E+00&       1.23E+00\\ 
         1187.67&    3&    9&  3.60E+00&       1.38E+00&       1.24E+00&       1.11E+00&       9.98E-01&       9.18E-01\\ 
          804.69&    1&    8&  9.31E-02&       9.05E-01&       8.10E-01&       7.23E-01&       6.47E-01&       5.94E-01\\ 
         1662.71&    5&   10&  6.07E-01&       4.79E+00&       4.26E+00&       3.77E+00&       3.35E+00&       3.05E+00\\ 
         1147.80&    2&    7&  3.00E-02&       1.51E+00&       1.36E+00&       1.22E+00&       1.10E+00&       1.02E+00\\
        \hline
    \end{tabular}
	\caption{Effective collision strengths for W~{\sc iii}.}
	\label{tab:eiew2}
\end{table*}

%% file: eietablePt0.tex
\begin{table*}
	\begin{tabular}{ccccccccc}
		\hline
Wavelength      &\multicolumn{2}{c}{Index}& A-value & \multicolumn{5}{c}{Effective Collision Strengths} \\ 
(nm) & Lower&Upper & (s$^{-1}$) & 0.15~eV & 0.25~eV & 0.4~eV & 0.6~eV & 0.8~eV \\
\hline
         1522.66&    1&    5&  5.25E-01&       2.69E+00&       2.51E+00&       2.34E+00&       2.19E+00&       2.09E+00\\ 
        12888.66&    1&    2&  4.09E-04&       5.16E+00&       4.88E+00&       4.61E+00&       4.38E+00&       4.22E+00\\ 
         1076.07&    3&    6&  2.52E+00&       9.62E-01&       9.14E-01&       8.70E-01&       8.32E-01&       8.05E-01\\ 
         1726.65&    2&    5&  1.40E-02&       2.35E+00&       2.19E+00&       2.05E+00&       1.93E+00&       1.84E+00\\ 
         1068.83&    2&    7&  1.16E+00&       8.73E-01&       7.77E-01&       6.89E-01&       6.13E-01&       5.59E-01\\ 
         1864.18&    2&    4&  4.36E-04&       8.20E-01&       7.40E-01&       6.66E-01&       6.02E-01&       5.57E-01\\ 
          740.95&    1&    8&  9.59E-01&       1.91E+00&       1.80E+00&       1.69E+00&       1.59E+00&       1.53E+00\\ 
         1070.57&    2&    6&  1.22E-01&       1.30E+00&       1.26E+00&       1.22E+00&       1.18E+00&       1.16E+00\\ 
         2805.51&    5&    7&  1.32E-01&       1.50E+00&       1.41E+00&       1.33E+00&       1.26E+00&       1.21E+00\\ 
         1741.01&    3&    5&  7.74E-05&       2.02E+00&       1.81E+00&       1.61E+00&       1.44E+00&       1.32E+00\\ 
        \hline
    \end{tabular}
	\caption{Effective collision strengths for Pt~{\sc i}.}
	\label{tab:eiept0}
\end{table*}

%% file: eietablePt1.tex
\begin{table*}
	\begin{tabular}{ccccccccc}
		\hline
Wavelength      &\multicolumn{2}{c}{Index}& A-value & \multicolumn{5}{c}{Effective Collision Strengths} \\ 
(nm) & Lower&Upper & (s$^{-1}$) & 0.15~eV & 0.25~eV & 0.4~eV & 0.6~eV & 0.8~eV \\
\hline
         2188.35&    2&    4&  2.54E+00&       1.35E+01&       1.24E+01&       1.14E+01&       1.05E+01&       9.84E+00\\ 
         1187.67&    1&    3&  9.05E+00&       1.82E+00&       1.71E+00&       1.61E+00&       1.52E+00&       1.46E+00\\ 
         1068.80&    1&    4&  1.10E-01&       2.39E+00&       2.28E+00&       2.17E+00&       2.07E+00&       2.01E+00\\ 
         2517.02&    4&    5&  1.96E+00&       4.49E+00&       4.11E+00&       3.76E+00&       3.47E+00&       3.25E+00\\ 
          633.26&    1&    6&  3.85E+00&       1.41E+00&       1.33E+00&       1.25E+00&       1.19E+00&       1.14E+00\\ 
         2036.89&    3&    5&  2.68E-01&       2.02E+00&       1.86E+00&       1.71E+00&       1.58E+00&       1.49E+00\\ 
         2089.14&    1&    2&  6.17E-06&       3.07E+00&       2.96E+00&       2.86E+00&       2.77E+00&       2.71E+00\\ 
         1339.65&    4&    7&  6.46E+00&       2.88E+00&       2.65E+00&       2.44E+00&       2.26E+00&       2.13E+00\\ 
          751.25&    2&    8&  1.52E+01&       3.83E+00&       3.46E+00&       3.13E+00&       2.85E+00&       2.64E+00\\ 
          750.23&    1&    5&  3.93E-02&       2.16E+00&       2.05E+00&       1.95E+00&       1.86E+00&       1.80E+00\\ 
        \hline
    \end{tabular}
	\caption{Effective collision strengths for Pt~{\sc ii}.}
	\label{tab:eiept1}
\end{table*}

%% file: eietablePt2.tex
\begin{table*}
	\begin{tabular}{ccccccccc}
		\hline
Wavelength      &\multicolumn{2}{c}{Index}& A-value & \multicolumn{5}{c}{Effective Collision Strengths} \\ 
(nm) & Lower&Upper & (s$^{-1}$) & 0.15~eV & 0.25~eV & 0.4~eV & 0.6~eV & 0.8~eV \\
\hline
         1025.46&    1&    3&  4.66E+01&       4.27E+00&       4.11E+00&       3.97E+00&       3.85E+00&       3.76E+00\\ 
         1126.28&    2&    4&  2.22E+01&       2.83E+00&       2.60E+00&       2.39E+00&       2.21E+00&       2.08E+00\\ 
         2242.86&    2&    3&  1.13E+00&       3.82E+00&       3.56E+00&       3.33E+00&       3.12E+00&       2.98E+00\\ 
          870.44&    2&    6&  2.10E+01&       1.61E+00&       1.47E+00&       1.34E+00&       1.23E+00&       1.15E+00\\ 
         2262.34&    3&    4&  3.05E+00&       3.73E+00&       3.43E+00&       3.16E+00&       2.93E+00&       2.76E+00\\ 
          468.81&    1&    7&  1.91E+01&       2.68E+00&       2.65E+00&       2.62E+00&       2.60E+00&       2.58E+00\\ 
         3831.85&    4&    6&  4.05E-01&       2.12E+00&       1.99E+00&       1.88E+00&       1.78E+00&       1.71E+00\\ 
          863.63&    3&    7&  1.94E+00&       4.02E+00&       3.68E+00&       3.38E+00&       3.12E+00&       2.93E+00\\ 
          457.94&    1&    8&  1.02E-02&       3.11E+00&       3.07E+00&       3.04E+00&       3.01E+00&       2.99E+00\\ 
         5427.12&    5&    6&  2.00E-01&       5.85E-01&       5.37E-01&       4.92E-01&       4.53E-01&       4.26E-01\\ 
        \hline
    \end{tabular}
	\caption{Effective collision strengths for Pt~{\sc iii}.}
	\label{tab:eiept2}
\end{table*}

%% file: eietableAu0.tex
\begin{table*}
	\begin{tabular}{ccccccccc}
		\hline
Wavelength      &\multicolumn{2}{c}{Index}& A-value & \multicolumn{5}{c}{Effective Collision Strengths} \\ 
(nm) & Lower&Upper & (s$^{-1}$) & 0.15~eV & 0.25~eV & 0.4~eV & 0.6~eV & 0.8~eV \\
\hline
         1091.56&    1&    2&  2.48E-02&       1.55E+00&       1.49E+00&       1.45E+00&       1.41E+00&       1.38E+00\\ 
          814.73&    2&    3&  2.98E+01&       6.45E-01&       6.06E-01&       5.70E-01&       5.39E-01&       5.17E-01\\ 
          466.52&    1&    3&  1.31E+00&       1.10E+00&       1.02E+00&       9.48E-01&       8.87E-01&       8.44E-01\\ 
          267.67&    1&    4&  1.61E+08&       8.30E-01&       9.85E-01&       1.13E+00&       1.25E+00&       1.34E+00\\ 
          242.87&    1&    5&  2.26E+08&       9.96E-01&       1.28E+00&       1.55E+00&       1.78E+00&       1.94E+00\\ 
          303.01&    2&    6&  1.37E+05&       2.34E+00&       2.33E+00&       2.31E+00&       2.30E+00&       2.29E+00\\ 
          627.99&    3&    4&  2.33E+06&       4.18E-02&       1.32E-01&       2.16E-01&       2.88E-01&       3.39E-01\\ 
          312.37&    2&    5&  2.60E+07&       2.60E-01&       3.50E-01&       4.34E-01&       5.06E-01&       5.57E-01\\ 
          274.91&    2&    7&  1.43E+06&       4.47E+00&       4.10E+00&       3.76E+00&       3.47E+00&       3.26E+00\\ 
          268.69&    2&    9&  1.21E+06&       2.10E+00&       2.11E+00&       2.13E+00&       2.14E+00&       2.15E+00\\ 
        \hline
    \end{tabular}
	\caption{Effective collision strengths for Au~{\sc i}.}
	\label{tab:eieau0}
\end{table*}

%% file: eietableAu1.tex
\begin{table*}
	\begin{tabular}{ccccccccc}
		\hline
Wavelength      &\multicolumn{2}{c}{Index}& A-value & \multicolumn{5}{c}{Effective Collision Strengths} \\ 
(nm) & Lower&Upper & (s$^{-1}$) & 0.15~eV & 0.25~eV & 0.4~eV & 0.6~eV & 0.8~eV \\
\hline
          566.87&    1&    3&  4.05E-01&       5.29E-01&       5.76E-01&       6.18E-01&       6.55E-01&       6.81E-01\\ 
         3844.62&    2&    3&  2.87E-01&       7.68E+00&       7.61E+00&       7.54E+00&       7.48E+00&       7.44E+00\\ 
          987.64&    3&    4&  2.72E+01&       2.00E+00&       1.78E+00&       1.57E+00&       1.40E+00&       1.27E+00\\ 
          685.79&    2&    5&  2.71E+01&       1.74E+00&       1.60E+00&       1.47E+00&       1.36E+00&       1.28E+00\\ 
          337.60&    1&    5&  8.27E+00&       9.88E-01&       9.16E-01&       8.50E-01&       7.92E-01&       7.52E-01\\ 
          834.68&    3&    5&  1.74E+00&       1.60E+00&       1.50E+00&       1.41E+00&       1.34E+00&       1.28E+00\\ 
          785.78&    2&    4&  4.98E-03&       7.74E-01&       6.96E-01&       6.24E-01&       5.62E-01&       5.18E-01\\ 
         5389.40&    4&    5&  5.42E-02&       5.07E+00&       4.94E+00&       4.83E+00&       4.73E+00&       4.66E+00\\ 
          393.09&    2&    6&  5.34E+00&       2.09E+00&       2.61E+00&       3.09E+00&       3.51E+00&       3.80E+00\\ 
          437.86&    3&    6&  7.39E-01&       1.61E+00&       1.74E+00&       1.85E+00&       1.95E+00&       2.02E+00\\ 
        \hline
    \end{tabular}
	\caption{Effective collision strengths for Au~{\sc ii}.}
	\label{tab:eieau1}
\end{table*}

%% file: eietableAu2.tex
\begin{table*}
	\begin{tabular}{ccccccccc}
		\hline
Wavelength      &\multicolumn{2}{c}{Index}& A-value & \multicolumn{5}{c}{Effective Collision Strengths} \\ 
(nm) & Lower&Upper & (s$^{-1}$) & 0.15~eV & 0.25~eV & 0.4~eV & 0.6~eV & 0.8~eV \\
\hline
          787.77&    1&    2&  3.30E+01&       3.39E+00&       3.36E+00&       3.33E+00&       3.31E+00&       3.29E+00\\ 
          336.09&    1&    3&  1.34E-01&       2.66E+00&       2.56E+00&       2.47E+00&       2.38E+00&       2.33E+00\\ 
          285.09&    1&    4&  1.05E+01&       2.39E+00&       2.29E+00&       2.20E+00&       2.11E+00&       2.06E+00\\ 
         1878.69&    3&    4&  3.72E+00&       1.17E+01&       1.10E+01&       1.03E+01&       9.73E+00&       9.32E+00\\ 
          257.58&    1&    5&  7.95E+00&       1.69E+00&       1.59E+00&       1.50E+00&       1.43E+00&       1.37E+00\\ 
          247.86&    1&    6&  2.04E+01&       9.73E-01&       9.49E-01&       9.27E-01&       9.07E-01&       8.94E-01\\ 
          382.73&    2&    5&  1.09E+00&       1.56E+00&       1.37E+00&       1.20E+00&       1.05E+00&       9.43E-01\\ 
          446.77&    2&    4&  2.78E-02&       1.27E+00&       1.17E+00&       1.08E+00&       1.01E+00&       9.52E-01\\ 
         2669.87&    4&    5&  1.01E+00&       5.10E+00&       4.90E+00&       4.73E+00&       4.58E+00&       4.47E+00\\ 
          225.09&    1&    7&  1.72E+01&       1.11E+00&       1.10E+00&       1.08E+00&       1.07E+00&       1.06E+00\\ 
        \hline
    \end{tabular}
	\caption{Effective collision strengths for Au~{\sc iii}.}
	\label{tab:eieau2}
\end{table*}